\documentclass[usenatbib,usegraphicx]{mn2e}
\usepackage{graphicx}
\usepackage{natbib}
\title[Si VI in DO white dwarf atmospheres]{Atomic data and electron-impact broadening effect in DO white
dwarf atmospheres: Si VI}
\author[R. Hamdi et al.]{R. Hamdi$^{1}$, N. Ben
Nessib$^{1}$, N. Milovanovi\'{c}$^{2}$, L.\v{C}.
Popovi\'{c}$^{2}$,
\newauthor
M.S. Dimitrijevi\'{c}$^{2}$\thanks{E-mail:
mdimitrijevic@aob.bg.ac.yu} and S. Sahal-Br\'{e}chot$^{3}$  \\
$^{1}$Groupe de Recherche en Physique Atomique et Astrophysique,
Institut National des Sciences\\ Appliqu\'{e}es et de Technologie,
Centre Urbain Nord B. P. No. 676, 1080 Tunis Cedex,
Tunisia.\\
$^{2}$Astronomical observatory, Volgina 7, 11160 Belgrade 74,
Serbia.\\ $^{3}$Laboratoire d'\'{E}tude du Rayonnement et de la
Mati\`{e}re en Astrophysique, UMR CNRS 8112,\\ Observatoire de
Paris-Meudon, 92195 Meudon, France.}

\begin{document}

\date{Accepted 1988 December 15. Received 1988 December 14; in original form 1988 October 11}

\pagerange{\pageref{firstpage}--\pageref{lastpage}} \pubyear{2008}

\maketitle

\label{firstpage}

\begin{abstract}
Energy levels, electric dipole transition probabilities and
oscillator strengths in five times ionized silicon have been
calculated in intermediate coupling. The present calculations were
carried out with the general purpose atomic-structure program
SUPERSTRUCTURE. The relativistic corrections to the
non-relativistic Hamiltonian are taken into account through the
Breit-Pauli approximation. We have also introduced a
semi-empirical correction (TEC) for the calculation of the
energy-levels. These atomic data are used to provide semiclassical
electron-, proton- and ionized helium- impact line widths and
shifts for 15 Si VI muliplet. Calculated results have been used to
consider the influence of Stark broadening for DO white dwarf
atmospheric conditions.
\end{abstract}


\section{Introduction}

Atomic data such as transition probabilities (A) play an important
role in the diagnostics and modelling of laboratory plasmas
\citep{b1}. Various kinetic processes appearing in plasma
modelling need reliable knowledge of A values. Further, knowledge
of A values gives a possibility for determination of coefficients
(B) which characterize the absorption and stimulated emission.
These processes are also important in laser physics. The
classification of transitions and determination of energy levels
are essential parts of the study of a laboratory spectrum. The
lack of available atomic data limits our ability to infer reliably
the properties of many cosmic plasmas and, hence, address many of
the fundamental issues in astrophysics \citep{b3}.

Accurate Stark broadening parameters are important to obtain a
reliable modelisation of stellar interiors. The Stark broadening
mechanism is also important for the investigation, analysis, and
modelling of B-type, and particulary A-type, stellar atmospheres,
as well as for white dwarf atmospheres \citep[see e.g.][]{b2,b4}.

Silicon, in various ionization stages, is detected in the
atmospheres of DO white dwarfs \citep{b5}. Si VI lines have been
observed as well for example in coronal line regions of planetary
nebulae NGC 6302 and 6537 \citep{b11}.

Uzelac and collaborators \citep{b6} studied plasma broadening of
Ne II-Ne VI and F IV-F V experimentally and theoretically, they
found that, the results of simplified semiclassical \citep[][Eq
526]{b1} calculations show better agreement at higher ionization
stages, while the modified semiempirical formula \citep{b7} seems
to be better for the low ionization stages. Unfortunately, due to
the lack of atomic data, most of the reported sophisticated
semiclassical Stark broadening parameters relate to spectral lines
of neutral and low ionization stages. In previous papers
\citep{b8,b9}, we calculated Stark broadening parameters of
quadruply ionized silicon and neon using SUPERSTRUCTURE and
\citet{b10} method for oscillator strengths and we found that the
difference is tolerable.

Si VI ion belongs to the fluorine-like sequence, its ground state
configuration is 1s$^{2}$2s$^{2}$2p$^{5}$ with the term
$^{2}$P${^{\circ }}$. In this work we present fine-structure
energy-levels, transition probabilities and oscillator strengths
for Si VI ion. The atomic structure code SUPERSTRUCTURE was used,
which allows for configuration interaction, relativistic effects
and semi-empirical term energy corrections. Calculated energies
and oscillator strengths are used to provide Stark broadening
parameters due to electron, proton, and ionized helium-impact of
Si VI lines. The obtained Stark broadening parameters are used to
investigate the influence of Stark broadening mechanism in hot,
high gravity star atmospheres as for example DO white dwarfs.

\section[]{The method}

In this work, the calculations were carried out with the
general-purpose atomic-structure program SUPERSTRUCTURE
\citep{b12}, as modified by \citet{b13}. The atomic model used to
calculate energies of terms or levels and transition probabilities include 26 \ configurations: 2s$^{2}$2p$^{5}$, 2s2p$%
^{6}$, 2s$^{2}$2p$^{4}$3$\ell$, 2s$^{2}$2p$^{4}$4$\ell$,
2s$^{2}$2p$^{4}$5$\ell$, 2s$^{2}$2p$^{4}$6$\ell$, 2s2p$^{5}
$3$\ell$ and 2p$^{6}$3$\ell$ ($\ell \leq n-1$). Configuration
interaction (CI) effects were fully taken into account. The wave
functions are of the type
\begin{equation}
\Psi =\sum\limits_{i}\Phi _{i}C_{i}
\end{equation}
where the basis functions $\Phi _{i}$ are constructed using
one-electron orbitals. The latter are calculated with a scaled
Thomas-Fermi statistical model potential or obtained from the
Coulomb potential. For each radial orbital $P_{nl}(r)$, the
potential can be adjusted using a parameter called $\lambda $. In
the present case, those {\it n} and {\it l}-dependent scaling
parameters
$\lambda _{nl}$ were determined variationally by optimizing the weighted sum of the term energies. The $P_{nl}$ are orthogonalised to each other such that the function  $%
P_{n_{1}l}$ is orthogonalised to the function $P_{n_{2}l}$ when $n_{2}<n_{1}$%
. The values adopted for the $\lambda _{nl}$ parameters are
presented in Table A. In this approach the Hamiltonian is taken to
be in the form
\begin{equation}
H=H_{nr}+H_{BP}
\end{equation}
relativistic corrections are included in Breit-Pauli Hamiltonian
($H_{BP}$) as perturbation to the non relativistic Hamiltonian
($H_{nr}$). $H_{BP}$ contains the one electron operators for the
mass correction, the Darwin contact term, the spin-orbit
interaction in the field of the nucleus and the two electron
operators for spin-orbit, spin-other orbit, and spin-spin
interactions. We also use the so-called term energy corrections
(TEC) introduced by \citet{b14}, in which the Hamiltonian matrix
is empirically adjusted to give the best agreement between
experimental energies and the final calculated term energies
including the relativistic effects. In practice, the TEC for a
given term is simply the difference between the calculated and
measured energy of the lowest level in the multiplet.

\bigskip

\noindent {\bf Table A.} {\small Optimized parameters $\lambda
_{nl}$ adopted for the 26 configuration model calculation. Positif
values for Thomas-Fermi-Dirac potential and negative values for
Coulomb potential.}
\begin{tabular}{llllllll}
\hline $n,l$ & $\lambda _{nl}$ & $n,l$ & $\lambda _{nl}$ & $n,l$ &
$\lambda _{nl}$ & $n,l$ & $\lambda _{nl}$ \\ \hline
1s & 1.4653 & 4s & 1.1091 & 5d & 1.1015 & 6f & -0.6629 \\
2s & 1.1844 & 4p & 1.0960 & 5f & 1.1135 & 6g & -1.0777 \\
2p & 1.1389 & 4d & 1.1103 & 5g & 1.0776 & 6h & -1.3002 \\
3s & 1.1078 & 4f & 1.0478 & 6s & 4.8162 &  &  \\
3p & 1.0762 & 5s & 1.1580 & 6p & 6.2669 &  &  \\
3d & 1.1808 & 5p & 1.0902 & 6d & 5.3865 &  &  \\ \hline
\end{tabular}

\bigskip

Stark broadening parameter calculations have been performed within
the semiclassical perturbation method \citep{b15,b16}. A detailed
description of this formalism with all the innovations is given in
\citet{b15,b16,b17,b18,b19,b20,b21}. The full halfwidth ($w$) and
shift ($d$) of an electron-impact broadened spectral line can be
expressed as:
\begin{eqnarray*}
W &=&N\int vf(v)dv\left( \sum\limits_{i^{\prime }\neq i}\sigma
_{ii^{\prime }}(v)+\sum\limits_{f^{\prime }\neq f}\sigma
_{ff^{\prime }}(v)+\sigma
_{el}\right) +W_{R} \\
\end{eqnarray*}
\begin{eqnarray}
d &=&N\int vf(v)dv\int\nolimits_{R_{3}}^{R_{D}}2\pi \rho d\rho
\sin (2\varphi _{p})
\end{eqnarray}
where $N\ $is the electron density, $f(\upsilon )\ $the Maxwellian
velocity distribution function for electrons, $\rho \ $denotes the
impact parameter of the incoming electron, $i\ $and$\ f\ $denote the
initial and the final atomic energy levels and $i^{\prime }\ $and$\
f\ ^{\prime }$ their corresponding perturbing levels, while $W_{R}\
$gives the contribution of the
Feshbach resonances \citep{b19}. The inelastic cross section $%
\sigma _{ii^{\prime }}(\upsilon )$ can be expressed by an integral
over the impact parameter of the transition probability
$P_{jj^{\prime }}(\rho ,\upsilon )\ $as
\begin{equation}
\sum_{j\neq j^{\prime }}\sigma _{ii^{\prime }}(\upsilon
)=\frac{1}{2}\pi R_{1}^{2}+\int_{R_{1}}^{R_{D}}2\pi \rho d\rho
\sum_{j\neq j^{\prime }}P_{jj^{\prime }}(\rho ,\upsilon ),\ j=i,f
\end{equation}
and the elastic cross section is given by
\begin{eqnarray*}
\sigma _{el}=2\pi R_{2}^{2}+\int_{R_{2}}^{R_{D}}2\pi \rho d\rho
\sin ^{2}\delta ,
\end{eqnarray*}
\begin{equation}
\delta =(\varphi _{p}^{2}+\varphi _{q}^{2})^{\frac{1}{2}}.
\end{equation}
The phase shifts $\varphi _{p}\ $and$\ \varphi _{q}\ \ $due
respectively to the polarization potential ($r^{-4}$) and to the quadrupolar potential ($%
r^{-3}$), are given in Section 3 of Chapter 2 in \citet{b15} and
$R_{D}$ is the Debye radius. All the cut-offs $R_{1},\ R_{2},\
R_{3}$ are described in Section 1 of Chapter 3 in \citet{b16}.

For electrons, hyperbolic paths due to the attractive Coulomb
force were used, while for perturbing ions the hyperbolic paths
are different since the force is repulsive. The formulae for the
ion-impact widths and shifts are analogous to Eqs. (3)-(5),
without the resonance contribution to the width.

\section{Results and discussion}

The calculated {\it ab initio} energies for Si VI are listed in
Table 1 along with experimentally determined energies for a number
of levels taken from a National Institute of Standards and
Technology (NIST) compilation. The configurations
for which we present results are 2s$^{2}$2p$^{5}$, 2s2p$^{6}$, 2s$^{2}$2p$%
^{4}$3$\ell$ , 2s$^{2}$2p$^{4}$4$\ell$$^{\prime }$. $\ell$ = s, p,
d and $\ell$$^{\prime }$ = s, p, d. We use the LS coupling scheme
to designate excited energy.

Two different models are used for the determination of energy
levels, the first contains the 9 first configurations of the model
given in Sect. 2 and the second contains the totality of the 26
configurations. Both 9-configuration model and more elaborated
26-configuration one give energy levels in good agreement with the
NIST values, indeed, our energies are lower than the NIST ones by
less than 1\% except for the
two first excited levels i.e. 2p$^{5}$\ $^{2}$P$%
{{}^\circ}%
_{1/2}$ and 2s2p$^{6}$\ $^{2}$S$_{1/2}$. But the results obtained
by the second model are always more accurate, that proves the
importance of the configuration interaction. Besides, if a term is
simply shifted relatively to the ground state, then the difference
with observed energy should be essentially constant. In some terms
the levels are not always in correct order. For example, the
observed order of the levels of 2p$^{4}$($^{3}$P)3d\ $^{4}$F is
(9/2, 7/2, 5/2, 3/2) and the present order is (9/2, 3/2, 5/2,
7/2).

We use the calculated energies and the wavefunctions to calculate
oscillator strengths and transition probabilities. With the aim of
improving the quality of our wavefunctions, the 2s2p$^{6}$\
$^{2}$S$_{1/2}$ level is corrected using TEC procedure
(see Sect. 2). This method can not be applied for 2p$^{5}$\ $^{2}$P$%
{{}^\circ}%
_{1/2}$ level. Electric dipole transition probabilities and weighted
oscillator strengths are presented in Table 2 for transition with
lower level from 1 to 10 and upper level from 3 to 95. The majority
of wavelengths are in XUV region. F-like ions are of fundamental
importance for current x-ray laser research. Our transition
probabilities are compared with NIST values and multiconfiguration
Hartree-Fock (MCHF) results of \citet{b22}, who use the observed
energies for the calculation of transition probabilities. The
agreement is by less than 30\% for 70\% of strong transitions (A $>$
10$^8$ s$^{-1}$) of NIST compilation. The agreement is much less for
weak ones. For weighted oscillator strengths, comparison is made
also with \citet{b23}results obtained  using multiconfiguration
Hartree-Fock relativistic (HFR) approach. In their work the adjusted
energy levels were used to optimize the electrostatic parameters,
these optimized parameters were used again to calculate the
gf-values. In general, our oscillator strengths are in good
agreement with the other works except for a few transitions as for
example 2-5, 1-6, 2-6, 2-38, for which we observe large
disagreement. For the transitions 1-47, 2-48, 2-42 our oscillator
strengths agree better with Coutinho \& Trigueiros results. On the
other hand, for the transitions 2-47, 1-48, 1-56 the agreement with
Froese-Fischer \& Tachiev values is better. For the transitions
3-15, 1-38, all three methods give different results.  The inclusion
of a larger number of configurations has an important effect on
wavefunctions and A values. TEC correction of the 2s2p$^{6}$\
$^{2}$S$_{1/2}$ level improve slightly our results. For example for
the 2-3 transition our A-value was 1.022E+10 s$^{-1}$ and became
8.840E+09 s$^{-1}$, NIST one is 8.46E+09 s$^{-1}$. The correction of
quartet terms do not improve the results.

By combining the SUPERSTRUCTURE code for calculating energy levels
and oscillator strengths and the code for the Stark broadening
calculations, we calculated Stark broadening parameters.
Calculated Stark broadening widths (FWHM) and shifts for a
perturber densities of 10$^{17}$ cm$^{-3}$ and temperatures from
50,000 up to 800,000 K are shown in Table 3 for electron-,
proton-, and singly ionized helium impact broadening. Such
temperatures are of interest for the modelling and analysis of
x-ray spectra, such as the spectra obtained by {\it Chandra},
modelling of some hot star atmospheres (e.g. DO white dwarf and PG
1195), subphotospheric layres, soft x-ray lasers and laser
produced plasmas. Higher temperatures are of interest for fusion
plasma as well as for stellar interiors.

We also specify a parameter \emph{C} \citep{b24}, which gives an
estimate for the maximal perturber density for which the line may
be treated as isolated, when it is divided by the corresponding
full width at half maximum. For each value given in Table 3 the
collision volume \emph{V} multiplied by the perturber density
\emph{N} is much less than one and the impact approximation is
valid \citep{b15,b16}. When the impact approximation is not valid,
the ion broadening contribution may be estimated by using the
quasistatic approach \citep{b1,b18,b25}.

Unfortunately, no experimental data are yet available for the
Stark broadening parameters so that the comparison is made only
with Dimitrijevi\'{c}'s (1993) results obtained using the modified
semiempirical formula \citep{b7}. All our values are greater than
Dimitrijevi\'{c}'s ones. The ratio $\frac{w_{e}}{w_{MSE}}$ shows
in average an agreement within 56\%. Low disagreements are usually
found for resonance lines, for example for the spectral line
2p$^{5}\ ^{2}$P$ {{}^\circ} $-2p$^{4}$($^{3}$P)3s $^{2}$P
($\lambda$ = 100,2 \AA) the ratio $\frac{w_{e}}{w_{MSE}}$ is only
1.06 for $T$= 800,000 K.

\section{Stark broadening effect in white dwarf atmospheres}

White dwarfs are separated in two distinct spectroscopic
sequences, the DA and non-DA white dwarfs. The former ones display
a pure hydrogen (optical) spectrum. The second, helium-rich
sequence comprise DO ($T$$_{eff}$ $>$ 45,000 K), DB (11,000 $<$
$T$$_{eff}$ $<$ 30,000 K) and DC ($T$$_{eff}$ $<$ 11,000 K) white
dwarfs. At the highest effective temperatures the DOs are
connected to the helium, carbon and oxygen-rich PG 1159.

Silicon in various ionization stages is present DO white dwarf
atmospheres \citep{b5}. We used our results for Stark widths to
examine the importance of electron-impact broadening in
atmospheres of DO white dwarfs for a trace element like Si VI.
Model atmospheres were taken from \citet{b27}.

 In hot star atmospheres, besides electron-impact broadening (Stark
broadening) the important broadening mechanism is a Doppler
(Thermal) one as well as the broadening due to the turbulence and
stellar rotation. Other types of spectral line broadening, such as
van der Waals, resonance and natural broadening, are usually
negligible. For a Doppler-broadened spectral lines, the intensity
distribution is not Lorentzian as for electron impact broadening
but Gaussian, whose full half width of the spectral lines may be
determined by the equation \citep[see e.g.][]{b28}
\begin{equation}
w_{D}[\AA]=7.16\times 10^{-7}\lambda \lbrack
\AA]\sqrt{\frac{T[K]}{M_{{Si}}}}
\end{equation}
where atomic weight for silicon is $M_{Si}$=28.1 a.u.

 The importance of Stark broadening in stellar atmospheres
is illustrated in Figs 1-4. In Fig 1 Stark (FWHM) and Doppler
widths for Si VI 2p$^{4}$($^{3}$P)3s $^{2}$P-2p$^{4}$($^{3}$P)3p
$^{2}$D$ {{}^\circ}$ ($\lambda$ = 1226,7 \AA) spectral line as a
function of atmospheric layer temperatures are shown. Stark widths
are shown for 6 atmospheric models with effective temperature
$T_{eff}$= 50,000 K-100,000 K and logarithm of surface gravity log
$g$=8. We can see in Fig 1. that Stark broadening is more
important than Doppler broadening for deeper atmospheric layers
for all effective temperatures. For WD with effective temperature
$T_{eff}$= 50,000 K, Stark and Doppler widths are equal for
temperature layer $T$$\approx$ 70,000 K and for WD with effective
temperature $T_{eff}$= 100,000 K, Stark and Doppler are equal for
temperature layer $T$$\approx$ 125,000 K. One should take into
account however, that even when the Doppler width is larger than
Stark width, due to different behaviour of Gaussian and Lorentzian
distributions, Stark broadening may be important in line wings. In
Fig 2. we present Stark (FWHM) and Doppler widths for Si VI
($\lambda$ = 1226,7 \AA) spectral line as a function of Rosseland
optical depth for the same atmospheric models as in Fig 1.

 The dependence of Stark and Doppler broadening in atmospheric
layer temperature for 4 values of surface gravity is shown in Fig
3. Atmospheric models used here have effective temperature
$T_{eff}$= 80,000 K. For stellar atmosphere with higher values of
surface gravity (log $g$ = 8-9), Stark broadening is significantly
larger than Doppler one. For stellar atmosphere with surface
gravity log $g$ = 7, Stark widths are comparable to Doppler widths
only for deeper hot atmospheric layer. For stellar atmospheres
with log $g$ = 6, Doppler broadening is dominant for all
atmospheric layers.

\section{Conclusions}

In present work we have calculated $ab$ $initio$ energy levels for
the eight lowest configurations of Si VI. We have also calculated
transition probabilities and oscillator strengths for 288
transitions. These data are useful for interpretation of laboratory
and astrophysical spectra, since, the reliability of the predicted
emergent spectra and the derived spectral diagnoses is directly
influenced by the quality of radiative data. The method used here is
semirelativistic one, the relativistic corrections are included by
using the Breit-Pauli Hamiltonian as perturbation to the non
relativistic Hamiltonian. To make fully relativistic calculation,
the GRASP code \citep{b29} can be used. One should note also that
\citet{MW76}investigated the influence of relativistic effects on
the oscillator strength values for the lithium isoelectronic
sequences and found that the influence is not important on
investigated f values for the ionization degrees investigated in our
work. We have reported results of Stark broadening parameter
calculations for 15 spectral lines of Si VI. For the simple
spectrum, the Stark broadening parameters of different lines are
nearly the same within a multiplet \citep{b30}. Consequently, we
have used the averaged atomic data for a multiplet as a whole and
calculate the corresponding Stark widths and shifts. We see that
using the SUPERSTRUCTURE code one obtains a set of energy levels and
oscillator strengths, enabling a calculation of Stark broadening
parameters when other theoretical and experimental data do not
exist. The Stark broadening parameters obtained here, contribute to
the creation of a set of such data for as large as possible number
of spectral lines, of significance for a number of problems in
astrophysical, laboratory and technological plasma research. Our
analysis of the influence of Stark broadening on Si VI ($\lambda$ =
1226,7 \AA) spectral line for stellar plasma conditions,
demonstrates the importance of this broadening mechanism for hot,
high gravity star atmospheres as for example DO white dwarfs.

\section*{Acknowledgments}
We would like to thank C.J. Zeippen for providing his version of
SUPERSTRUCTURE code. This work is a part of the projects 146001
"Influence of collisional processes on astrophysical plasma line
shapes" and 146002 "Astrophysical Spectroscopy of Extragalactic
Objects" supported by the Ministry of Science of Serbia.

\begin{figure}
{\includegraphics[width=8.5cm]{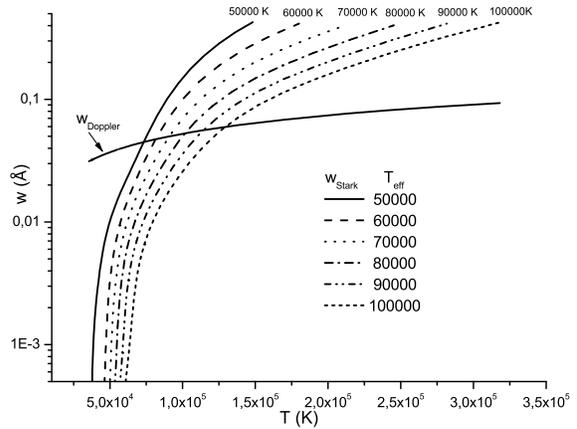}}
\caption{Stark and Doppler widths for Si VI 2p$^{4}$($^{3}$P)3s
$^{2}$P-2p$^{4}$($^{3}$P)3p $^{2}$D$ {{}^\circ}$ ($\lambda
$=1226,7 \AA) spectral line as a function of atmospheric layer
temperatures. Stark widths are shown for 6 atmospheric models with
effective temperature $T_{eff}$= 50,000 K-100,000 K and log
$g$=8.}
\end{figure}
\begin{figure}
 {\includegraphics[width=8.5cm]{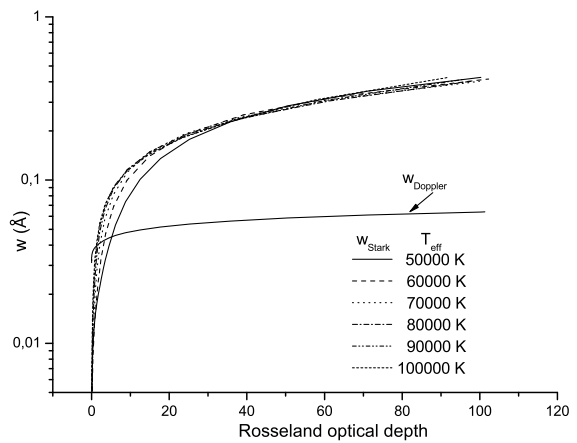}}
  \caption{Stark and Doppler widths for Si VI
2p$^{4}$($^{3}$P)3s $^{2}$P-2p$^{4}$($^{3}$P)3p $^{2}$D$
{{}^\circ}$ ($\lambda $=1226,7 \AA) spectral line as a function of
Rosseland optical depth. Stark widths are shown for 6 atmospheric
models with effective temperature $T_{eff}$= 50,000 K-100,000 K
and log $g$=8.}
\end{figure}
\begin{figure}
{\includegraphics[width=8.5cm]{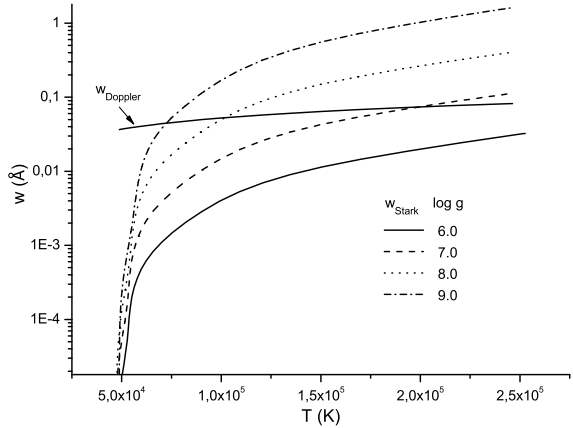}}
  \caption{Stark and Doppler widths for Si VI
2p$^{4}$($^{3}$P)3s $^{2}$P-2p$^{4}$($^{3}$P)3p $^{2}$D$
{{}^\circ}$ ($\lambda $=1226,7 \AA) spectral line as a function of
atmospheric layer temperatures. Stark widths are shown for 4
values of model gravity log $g$ = 6-9, $T_{eff}$= 80,000 K.}
\end{figure}
\begin{figure}
 {\includegraphics[width=8.5cm]{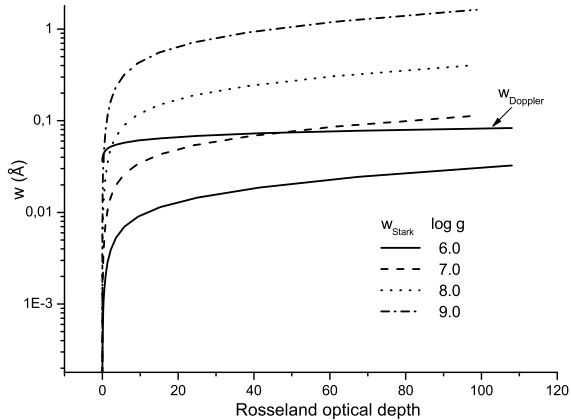}}
  \caption{Stark and Doppler widths for Si VI
2p$^{4}$($^{3}$P)3s $^{2}$P-2p$^{4}$($^{3}$P)3p $^{2}$D$
{{}^\circ}$ ($\lambda $=1226,7 \AA) spectral line as a function of
Rosseland optical depth. Stark widths are shown for 4 values of
model gravity log $g$ = 6-9, $T_{eff}$= 80,000 K.}
\end{figure}
\begin{table*}
\caption{Energy levels for Si VI in cm$^{-1}$. Key: a number assigned to each level. $%
LS$: $LS$ term and parity (superscript $%
{{}^\circ}%
\ $designate an odd level). $J$: \ $J$ value of the level.
E$_{9-conf}$: $ab$ $initio$ energy levels, calculated with
9-configuration model. E$_{26-conf}$: $ab$ $initio$ energy levels,
calculated with 26-configuration model. E$_{NIST}$: NIST values.}
\label{tab:1}       
\begin{center}
\begin{tabular}{llllrrr}
\hline $Key$ & $configuration$ & $LS$ & $J$&E$_{9-conf}$ & E$_{26-conf}$ & E$_{NIST}$ \\
\hline
$1$ & $2p^{5}$ & $^{2}P%
{{}^\circ}%
$ & $3/2$&0 & $0$ & $0$ \\
$2$ &  &  & $1/2$&2796 & $2839$ & $5090$ \\
$3$ & $2s2p^{6}$ & $^{2}S$ & $1/2$&435632 & $426462$ & $406497$ \\
$4$ & $2p^{4}(^{3}P)3s$ & $^{4}P$ & $5/2$&971851 & $982366$ & $990516$ \\
$5$ &  &  & $1/2$&975520 & $986496$ & $995470$ \\
$6$ &  &  & $3/2$&975969 & $987021$ & $993640$ \\
$7$ & $2p^{4}(^{3}P)3s$ & $^{2}P$ & $3/2$&987370 & $999127$ & $1005430$ \\
$8$ &  &  & $1/2$&987692 & $999494$ & $1009118$ \\
$9$ & $2p^{4}(^{1}D)3s$ & $^{2}D$ & $5/2$&1024374 & $1036129$ & $1041416$ \\
$10$ &  &  & $3/2$&1024884 & $1036533$ & $1041472$ \\
$11$ & $2p^{4}(^{3}P)3p$ & $^{4}P%
{{}^\circ}%
$ & $3/2$&1052168 & $1061210$ & $1069854$ \\
$12$ &  &  & $5/2$&1052404 & $1061401$ & $1068813$ \\
$13$ &  &  & $1/2$&1053583 & $1062874$ & $1071129$ \\
$14$ & $2p^{4}(^{3}P)3p$ & $^{4}D%
{{}^\circ}%
$ & $7/2$&1059089 & $1070621$ & $1078935$ \\
$15$ &  &  & $1/2$&1061265 & $1073242$ & $1083003$ \\
$16$ &  &  & $3/2$&1062447 & $1074609$ & $1082215$ \\
$17$ &  &  & $5/2$&1062788 & $1074942$ & $1080700$ \\
$18$ & $2p^{4}(^{3}P)3p$ & $^{2}D%
{{}^\circ}%
$ & $5/2$&1068139 & $1080203$ & $1086796$ \\
$19$ &  &  & $3/2$&1068477 & $1080547$ & $1089547$ \\
$20$ & $2p^{4}(^{3}P)3p$ & $^{2}P%
{{}^\circ}%
$ & $1/2$&1077264 & $1080958$ &  \\
$21$ &  &  & $3/2$&1075947 & $1084287$ & $1092171$ \\
$22$ & $2p^{4}(^{1}S)3s$ & $^{2}S$ & $1/2$&1071075 & $1084295$ & $1094449$ \\
$23$ & $2p^{4}(^{3}P)3p$ & $^{4}S%
{{}^\circ}%
$ & $3/2$&1074673 & $1087671$ & $1093752$ \\
$24$ & $2p^{4}(^{3}P)3p$ & $^{2}S%
{{}^\circ}%
$ & $1/2$&1094162 & $1088558$ &  \\
$25$ & $2p^{4}(^{1}D)3p$ & $^{2}F%
{{}^\circ}%
$ & $5/2$&1106342 & $1118373$ & $1123540$ \\
$26$ &  &  & $7/2$&1106472 & $1118689$ & $1124219$ \\
$27$ & $2p^{4}(^{1}D)3p$ & $^{2}D%
{{}^\circ}%
$ & $3/2$&1116033 & $1128816$ & $1134081$ \\
$28$ &  &  & $5/2$&1116641 & $1129460$ & $1134496$ \\
$29$ & $2p^{4}(^{1}D)3p$ & $^{2}P%
{{}^\circ}%
$ & $3/2$&1140823 & $1145932$ & $1147901$ \\
$30$ &  &  & $1/2$&1140921 & $1145980$ & $1150282$ \\
$31$ & $2p^{4}(^{1}S)3p$ & $^{2}P%
{{}^\circ}%
$ & $3/2$&1182143 & $1171561$ &  \\
$32$ &  &  & $1/2$&1182942 & $1172417$ &  \\
$33$ & $2p^{4}(^{3}P)3d$ & $^{4}D$ & $7/2$&1161013 & $1172209$ & $1181167$ \\
$34$ &  &  & $5/2$&1162025 & $1173271$ & $1181649$ \\
$35$ &  &  & $3/2$&1162165 & $1173376$ & $1182311$ \\
$36$ &  &  & $1/2$&1162754 & $1173979$ & $1182894$ \\
$37$ & $2p^{4}(^{3}P)3d$ & $^{4}F$ & $9/2$&1169569 & $1182612$ & $1189844$ \\
$38$ &  &  & $3/2$&1171689 & $1184525$ & $1194327$ \\
$39$ &  &  & $5/2$&1172030 & $1184838$ & $1193223$ \\
$40$ &  &  & $7/2$&1172047 & $1184946$ & $1191541$ \\
$41$ & $2p^{4}(^{3}P)3d$ & $^{4}P$ & $1/2$&1176385 & $1188594$ & $1194899$ \\
$42$ &  &  & $3/2$&1177207 & $1189543$ & $1195984$ \\
$43$ &  &  & $5/2$&1178908 & $1191510$ & $1197727$ \\
$44$ & $2p^{4}(^{3}P)3d$ & $^{2}F$ & $7/2$&1177148 & $1189950$ & $1194987$ \\
$45$ &  &  & $5/2$&1178364 & $1190848$ & $1197151$ \\
$46$ & $2p^{4}(^{3}P)3d$ & $^{2}P$ & $1/2$&1182736 & $1194144$ & $1200710$ \\
$47$ &  &  & $3/2$&1185454 & $1196939$ & $1204740$ \\
$48$ & $2p^{4}(^{3}P)3d$ & $^{2}D$ & $3/2$&1185167 & $1195061$ & $1201100$ \\
$49$ &  &  & $5/2$&1185973 & $1196247$ & $1202960$ \\
$50$ & $2p^{4}(^{1}D)3d$ & $^{2}G$ & $7/2$&1214568 & $1227449$ &  \\
$51$ &  &  & $9/2$&1214715 & $1227538$ & $1232671$ \\
$52$ & $2p^{4}(^{1}D)3d$ & $^{2}S$ & $1/2$&1223636 & $1234097$ & $1239190$ \\
$53$ & $2p^{4}(^{1}D)3d$ & $^{2}P$ & $3/2$&1224780 & $1234806$ & $1241050$ \\
$54$ &  &  & $1/2$&1225778 & $1235883$ & $1242390$ \\
$55$ & $2p^{4}(^{1}D)3d$ & $^{2}F$ & $7/2$&1223872 & $1236910$ & $1242649$ \\
$56$ &  &  & $5/2$&1223946 & $1236950$ & $1242186$ \\
$57$ & $2p^{4}(^{1}D)3d$ & $^{2}D$ & $5/2$&1226710 & $1237621$ & $1243020$ \\
$58$ &  &  & $3/2$&1227611 & $1238313$ & $1243860$ \\ \hline
\end{tabular}
\end{center}
\end{table*}
\newpage
\addtocounter{table}{-1}
\begin{table*}
\caption{$continued.$}
\label{tab:1}       
\begin{center}
\begin{tabular}{llllrrr}
\hline $Key$ & $configuration$ & $LS$ & $J$&E$_{9-conf}$ &
E$_{26-conf}$ & E$_{NIST}$ \\\hline
$59$ & $2p^{4}(^{1}S)3d$ & $^{2}D$ & $3/2$&1290487 & $1280925$ & $1291790$ \\
$60$ &  &  & $5/2$&1291154 & $1281640$ & $1291510$ \\
$61$ & $2p^{4}(^{3}P)4s$ & $^{4}P$ & $5/2$&1302866 & $1314745$ &  \\
$62$ &  &  & $1/2$&1305283 & $1317544$ &  \\
$63$ &  &  & $3/2$&1306890 & $1319305$ &  \\
$64$ & $2p^{4}(^{3}P)4s$ & $^{2}P$ & $3/2$&1308844 & $1321429$ & $1329900$ \\
$65$ &  &  & $1/2$&1310323 & $1323045$ &  \\
$66$ & $2p^{4}(^{3}P)4p$ & $^{4}P%
{{}^\circ}%
$ & $5/2$&1333771 & $1340871$ &  \\
$67$ &  &  & $3/2$&1333474 & $1340979$ &  \\
$68$ &  &  & $1/2$&1334815 & $1342729$ &  \\
$69$ & $2p^{4}(^{3}P)4p$ & $^{4}D%
{{}^\circ}%
$ & $7/2$&1335203 & $1347589$ &  \\
$70$ &  &  & $1/2$&1337989 & $1350092$ &  \\
$71$ &  &  & $3/2$&1338555 & $1351294$ &  \\
$72$ &  &  & $5/2$&1338545 & $1352144$ &  \\
$73$ & $2p^{4}(^{3}P)4p$ & $^{2}D%
{{}^\circ}%
$ & $5/2$&1339973 & $1351228$ &  \\
$74$ &  &  & $3/2$&1339638 & $1352195$ &  \\
$75$ & $2p^{4}(^{3}P)4p$ & $^{2}S%
{{}^\circ}%
$ & $1/2$&1340521 & $1352722$ &  \\
$76$ & $2p^{4}(^{3}P)4p$ & $^{4}S%
{{}^\circ}%
$ & $3/2$&1342995 & $1355616$ &  \\
$77$ & $2p^{4}(^{3}P)4p$ & $^{2}P%
{{}^\circ}%
$ & $3/2$&1349411 & $1359114$ &  \\
$78$ &  &  & $1/2$&1349310 & $1359465$ &  \\
$79$ & $2p^{4}(^{1}D)4s$ & $^{2}D$ & $5/2$ &1352262& $1365126$ & $1371820$ \\
$80$ &  &  & $3/2$&1352767 & $1365515$ &  \\
$81$ & $2p^{4}(^{3}P)4d$ & $^{4}D$ & $7/2$&1371226 & $1382797$ &  \\
$82$ &  &  & $5/2$&1372174 & $1384161$ &  \\
$83$ &  &  & $3/2$&1372430 & $1384614$ &  \\
$84$ &  &  & $1/2$&1373107 & $1385519$ &  \\
$85$ & $2p^{4}(^{3}P)4d$ & $^{4}F$ & $9/2$&1374444 & $1387108$ &  \\
$86$ &  &  & $7/2$&1376125 & $1388760$ &  \\
$87$ &  &  & $5/2$&1376654 & $1389343$ &  \\
$88$ &  &  & $3/2$&1376461 & $1389348$ &  \\
$89$ & $2p^{4}(^{3}P)4d$ & $^{4}P$ & $1/2$&1377931 & $1390631$ &  \\
$90$ &  &  & $3/2$&1379041 & $1391893$ &  \\
$91$ &  &  & $5/2$&1380830 & $1393884$ &  \\
$92$ & $2p^{4}(^{3}P)4d$ & $^{2}F$ & $5/2$&1380128 & $1392885$ &  \\
$93$ &  &  & $7/2$&1380477 & $1393669$ &  \\
$94$ & $2p^{4}(^{3}P)4d$ & $^{2}P$ & $1/2$&1381851 & $1393521$ & $1402490$ \\
$95$ &  &  & $3/2$&1384583 & $1396287$ & $1403050$ \\
$96$ & $2p^{4}(^{3}P)4d$ & $^{2}D$ & $3/2$&1385040 & $1395603$ &  \\
$97$ &  &  & $5/2$&1385544 & $1396797$ & $1404870$ \\
$98$ & $2p^{4}(^{3}P)4p$ & $^{2}F%
{{}^\circ}%
$ & $7/2$&1382852 & $1395696$ &  \\
$99$ &  &  & $5/2$&1382929 & $1395739$ &  \\
$100$ & $2p^{4}(^{1}D)4p$ & $^{2}D%
{{}^\circ}%
$ & $3/2$&1386127 & $1399126$ &  \\
$101$ &  &  & $5/2$&1386673 & $1399653$ &  \\
$102$ & $2p^{4}(^{1}D)4p$ & $^{2}P%
{{}^\circ}%
$ & $1/2$&1392385 & $1405752$ &  \\
$103$ &  &  & $3/2$&1392362 & $1405797$ &  \\
$104$ & $2p^{4}(^{1}D)4s$ & $^{2}S$ & $1/2$&1419675 & $1432175$ &  \\
$105$ & $2p^{4}(^{1}D)4d$ & $^{2}G$ & $7/2$&1421515 & $1434310$ &  \\
$106$ &  &  & $9/2$&1421617 & $1434434$ &  \\
$107$ & $2p^{4}(^{1}D)4d$ & $^{2}P$ & $3/2$&1424977 & $1436510$ &  \\
$108$ &  &  & $1/2$&1425973 & $1437711$ &  \\
$109$ & $2p^{4}(^{1}D)4d$ & $^{2}F$ & $7/2$&1424368 & $1437280$ &  \\
$110$ &  &  & $5/2$&1424453 & $1437401$ &  \\
$111$ & $2p^{4}(^{1}D)4d$ & $^{2}D$ & $5/2$&1425961 & $1438109$ & $1444340$ \\
$112$ &  &  & $3/2$&1426992 & $1438858$ & $1445010$ \\
$113$ & $2p^{4}(^{1}D)4d$ & $^{2}S$ & $1/2$&1427833 & $1440235$ &  \\
$114$ & $2p^{4}(^{1}S)4p$ & $^{2}P%
{{}^\circ}%
$ & $1/2$&1453543 & $1466060$ &  \\
$115$ &  &  & $3/2$&1453658 & $1466257$ &  \\
$116$ & $2p^{4}(^{1}S)4d$ & $^{2}D$ & $3/2$ &1493396& $1506902$ &  \\
$117$ &  &  & $5/2$&1494092 & $1507702$ &  \\ \hline
\end{tabular}
\end{center}
\end{table*}
\begin{table*}
\caption{Transition probabilities (A$_{ki}$), calculated
wavelengths ($\lambda $) and weighted oscillator strengths (gf)
for Si VI spectrum. present : this work, FF: \citet{b22}, CT:
\citet{b23}. The numbers in brackets denote powers of ten.}
\label{tab:2}       
\begin{center}
\begin{tabular}{lrllllllll}
\hline
$Transition$ & $\lambda ({%
{\AA}%
})$ &  $A_{ki}(s^{-1})$  &  &  &   &$gf$  &  \\
\cline{3-5}\cline{7-10} &   & $present$ & $FF$ & $NIST$ & &
$present$ & $FF$& $NIST$& $CT$ \\ \hline $1-3$ & $246.004$ &
$1.805(10)$ & $1.777(10)$ & $1.77(10)$ &
 & $3.275(-01)$&$3.210(-01)$ &$3.206(-01)$ & $4.04(-01)$ \\
$2-3$ & $247.734$  & $8.840(09)$ & $8.517(09)$ & $8.46(09)$ &
 & $1.627(-01)$ & $1.576(-01)$&$1.573(-01)$ & $2.00(-01)$ \\
$1-4$ & $101.795$ &  $8.982(07)$ & $7.405(07)$ & $7.51(07)$ &
 & $8.372(-04)$ & $6.784(-04)$&$6.886(-04)$ & $8.00(-04)$ \\
$1-5$ & $101.369$ & $1.340(09)$ & $8.423(06)$ & $8.55(06)$ &
 & $4.128(-03)$ & $2.547(-05)$&$2.588(-05)$ & \\
$2-5$ & $101.661$ & $2.010(09)$ & $2.624(08)$ & $2.64(08)$ &
 & $6.229(-03)$ & $8.016(-04)$&$8.072(-04)$ & $9.00(-04)$ \\
$1-6$ & $101.315$ & $1.019(08)$ & $1.167(09)$ & $1.18(09)$ &
 & $6.270(-04)$ & $7.088(-03)$&$7.161(-03)$ & $7.50(-03)$ \\
$2-6$ & $101.607$ & $9.716(06)$ & $1.010(08)$ & $1.08(08)$ &
 & $6.016(-05)$ & $6.194(-04)$&$6.266(-04)$ & $6.00(-04)$ \\
$1-7$ & $100.087$  & $6.667(10)$ & $6.753(10)$ & $6.74(10)$ &
 & $4.005(-01)$ & $4.004(-01)$&$3.999(-01)$ & $4.88(-01)$ \\
$2-7$ & $100.373$ & $1.292(10)$ & $1.098(10)$ & $1.09(10)$ &
 & $7.803(-02)$ & $6.574(-02)$&$6.561(-02)$ &$7.90(-02)$ \\
$1-8$ & $100.051$& $2.509(10)$ & $2.823(10)$ & $2.82(10)$ &
 & $7.530(-02)$ & $8.304(-02)$&$8.317(-02)$ & $1.02(-01)$ \\
$2-8$ & $100.336$ & $4.987(10)$ & $5.156(10)$ & $5.15(10)$ &
 & $1.505(-01)$ & $1.532(-01)$&$1.531(-01)$ & $1.86(-01)$ \\
$1-9$ & $96.513$ &  $3.189(10)$ & $3.087(10)$ & $3.08(10)$ &  &
 $2.672(-01)$ & $2.556(-01)$&$2.558(-01)$ &  \\
$1-10$ & $96.475$ & $4.625(09)$ & $3.056(09)$ & $3.03(09)$ &
 & $2.582(-02)$ & $1.687(-02)$&$1.674(-02)$ & $2.03(-02)$ \\
$2-10$ & $96.740$ & $2.674(10)$ & $2.796(10)$ & $2.79(10)$ &
 & $1.501(-01)$ & $1.558(-01)$&$1.559(-01)$ & $1.94(-01)$ \\
$3-11$ & $152.739$ & $1.064(05)$ & $7.895(05)$ & &
 & $1.488(-06)$ & $1.077(-05)$& &  \\
$4-11$ & $1268.336$ & $3.944(08)$ & $4.159(08)$ & $4.11(08)$ &  &
 $3.805(-01)$& $3.935(-01)$&$3.908(-01)$ &  \\
$5-11$ & $1338.445$ & $1.569(08)$ & $1.624(08)$ & $1.62(08)$ &  &
 $1.685(-01)$ & $1.746(-01)$& $1.749(-01)$& $1.89(-01)$ \\
$6-11$ & $1347.907$ & $8.664(07)$ & $9.855(07)$ & $9.73(07)$ &  &
 $9.440(-02)$ & $1.010(-01)$& $1.004(-01)$& $1.07(-01)$ \\
$7-11$ & $1610.745$ & $4.125(05)$ & $5.789(04)$ & $5.56(04)$ &  &
 $6.418(-04)$ & $8.312(-05)$&$8.035(-05)$ &  \\
$8-11$ & $1620.320$ &  $1.353(07)$ & $3.677(03)$ & $3.79(03)$ &  &
 $2.130(-02)$ & $5.938(-06)$&$6.165(-06)$ &  \\
$9-11$ & $3987.165$ &  $6.968(04)$ & $1.627(04)$ & $1.62(04)$ &  &
 $6.643(-04)$ & $1.218(-04)$& $1.199(-04)$&  \\
$10-11$ & $4052.352$ & $7.466(03)$ & $5.228(03)$ & $5.44(03)$ &  &
$7.353(-05)$ & $3.929(-05)$&$4.036(-05)$ &  \\
$4-12$ & $1265.271$  & $5.940(08)$ & $5.712(08)$ & $5.64(08)$ &  &
 $8.554(-01)$ & $8.322(-01)$& $8.279(-01)$& $8.72(-01)$ \\
$6-12$ & $1344.446$ &  $9.628(07)$ & $1.104(08)$ & $1.10(08)$ &  &
 $1.565(-01)$ & $1.744(-01)$&$1.753(-01)$ & $1.91(-01)$ \\
$7-12$ & $1605.805$ & $1.630(06)$ & $2.988(05)$ & $3.09(05)$ &  &
 $3.781(-03)$ & $6.648(-04)$&$6.918(-04)$ & \\
$9-12$ & $246.004$ & $1.458(05)$ & $2.110(03)$ & $2.37(03)$ &  &
 $2.054(-03)$ & $2.554(-05)$&$2.837(-05)$ &  \\
$10-12$ & $247.734$ & $3.719(01)$ & $2.179(02)$ & $2.40(02)$ &  &
 $5.410(-07)$ & $2.646(-06)$&$2.890(-06)$ &  \\
$3-13$ & $101.795$ & $2.939(05)$ & $4.160(05)$ & &
& $2.046(-06)$ & $2.628(-06)$& &  \\
$5-13$ & $1309.272$ &  $3.686(07)$ & $7.502(07)$ & $7.46(07)$ &  &
 $1.894(-02)$ & $3.900(-02)$& $3.899(-02)$&  \\
$6-13$ & $1318.325$ & $5.969(08)$ & $5.915(08)$ & $5.85(08)$ &  &
 $3.111(-01)$ & $2.933(-01)$& $2.917(-01)$& $3.10(-01)$ \\
$7-13$ & $1568.682$ & $1.079(06)$ & $8.921(05)$ & $9.49(05)$ &  &
 $7.964(-04)$ & $6.160(-04)$&$6.606(-04)$ &  \\
$8-13$ & $1577.761$ & $1.707(06)$ & $4.217(05)$ & $4.38(05)$ &  &
 $1.274(-03)$ & $3.268(-04)$&$3.419(-04)$ &  \\
$10-13$ & $3796.252$ & $1.573(04)$ & $1.490(04)$ & $1.49(04)$ &  &
 $6.798(-05)$ & $5.128(-05)$& $5.018(-05)$&  \\
$4-14$ & $1133.080$ & $1.056(09)$ & $1.033(09)$ & $1.03(09)$&  &
 $1.626(+00)$ & $1.581(+00)$&$1.577(+00)$ & $1.71(+00)$ \\
$9-14$ & $2899.222$ & $6.193(03)$ & $3.205(03)$ & $3.54(03)$ &  &
  $6.244(-05)$ & $2.776(-05)$&$3.019(-05)$ &  \\
$3-15$ & $149.982$ &  $5.880(04)$ & $1.367(06)$ & $1.40(06)$ &
& $3.966(-07)$ & $8.972(-06)$&$9.141(-06)$ & $9.00(-05)$ \\
$5-15$ & $1152.795$ & $9.112(08)$ & $8.776(08)$ & $8.74(08)$ &  &
 $3.631(-01)$ & $3.426(-01)$&$3.419(-01)$ & $3.70(-01)$ \\
$6-15$ & $1159.808$ & $6.171(07)$ & $1.290(08)$ & $1.29(08)$ &  &
 $2.489(-02)$ & $4.836(-02)$& $4.841(-02)$&  \\
$7-15$ & $1349.252$ & $9.550(05)$ & $2.566(05)$ & $2.58(05)$ &  &
 $5.213(-04)$ & $1.278(-04)$&$1.228(-04)$ &  \\
$8-15$ & $1355.964$ & $1.964(07)$ & $7.280(05)$ & $7.09(05)$ &  &
 $1.083(-02)$ & $3.998(-04)$& $3.899(-04)$& \\
$10-15$ & $2724.120$ & $6.668(04)$ & $8.238(03)$ &  &  &
 $1.484(-04)$ & $1.454(-05)$& &  \\
$3-16$ & $149.675$ & $3.954(05)$ & $5.543(05)$ & $5.73(05)$ &
& $5.312(-06)$ & $7.292(-06)$&$7.516(-06)$ &  \\
$4-16$ & $1084.091$ & $2.953(07)$ & $2.000(07)$ & $2.03(07)$ &  &
 $2.081(-02)$ & $1.423(-02)$&$1.452(-02)$ &  \\
$5-16$ & $1134.903$ & $4.668(08)$ & $4.842(08)$ & $4.82(08)$ &  &
 $3.606(-01)$ & $3.850(-01)$&$3.837(-01)$ & $4.13(-01)$ \\
$6-16$ & $1141.699$ & $5.093(08)$ & $4.957(08)$ &$4.95(08)$&  &
 $3.981(-01)$ & $3.781(-01)$&$3.768(-01)$ & $4.11(-01)$ \\
$7-16$ & $1324.806$ & $1.466(06)$ & $1.213(03)$ & $1.88(03)$ &  &
 $1.543(-03)$ & $1.232(-06)$&$1.909(-06)$ &  \\
$8-16$ & $1331.276$ & $2.832(07)$ & $4.541(06)$ & $4.53(06)$ &  &
 $3.010(-02)$ & $5.094(-03)$&$5.081(-03)$ &  \\
$9-16$ & $2598.741$ & $7.183(04)$ & $8.858(03)$ & $9.91(03)$ &  &
 $2.909(-04)$ & $3.241(-05)$& $3.572(-05)$& \\
$10-16$ & $2626.277$ & $1.035(02)$ & $1.337(04)$ & $1.45(04)$ &  &
 $4.281(-07)$ & $4.908(-05)$&$5.236(-05)$ & \\
$4-17$ & $1080.192$ & $2.103(08)$ & $1.903(08)$ & $1.91(08)$ &  &
 $2.207(-01)$ & $2.100(-01)$& $2.113(-01)$& $2.38(-01)$ \\
$6-17$ & $1137.375$ & $8.616(08)$ & $7.873(08)$ & $7.83(08)$ &  &
  $1.003(+00)$ & $9.320(-01)$&$9.289(-01)$ & $9.98(-01)$ \\
$7-17$ & $1318.987$ & $7.133(06)$ & $1.864(07)$ & $1.86(07)$ &  &
 $1.116(-02)$ & $2.957(-02)$&$2.958(-02)$ & \\
$9-17$ & $2576.446$ & $2.564(03)$ & $8.978(03)$ & $9.31(03)$ &  &
 $1.531(-05)$ & $5.316(-05)$&$5.432(-05)$ & \\
$4-18$ & $1022.110$ & $5.562(05)$ & $1.307(07)$ & $1.33(07)$ &  &
 $5.227(-04)$ & $1.265(-02)$&$1.285(-02)$ &  \\
$6-18$ & $1073.164$ & $1.749(07)$ & $2.238(07)$ & $2.22(07)$ &  &
 $1.811(-02)$ & $2.313(-02)$&$2.301(-02)$ &  \\ \hline
\end{tabular}
\end{center}
\end{table*}
\newpage
\addtocounter{table}{-1}
\begin{table*}
\caption{$Continued.$}
\label{tab:2}       
\begin{center}
\begin{tabular}{lrllllllll}
\hline
$Transition$ & $\lambda ({%
{\AA}%
})$  & $A_{ki}(s^{-1})$  &  &  &   &$gf$  &  \\
\cline{3-5}\cline{7-10} &   & $present$ & $FF$ & $NIST$ & &
$present$ & $FF$ &$NIST$ & $CT$ \\ \hline $7-18$ & $1233.405$  &
$8.506(08)$ & $8.032(08)$ & $8.02(08)$&  &
 $1.164(+00)$ & $1.090(-01)$&1.088(+00) &  \\
$9-18$ & $2268.921$  & $7.863(04)$ & $2.465(05)$ & $2.49(05)$ &  &
 $3.641(-04)$ & $1.090(-03)$& $1.086(-03)$& \\
$10-18$ & $2289.882$ & $1.355(02)$ & $6.522(03)$ & $6.85(03)$ &  &
 $6.391(-07)$ & $2.892(-05)$& $2.999(-05)$& \\
$3-19$ & $148.357$ &  $2.047(06)$ & $1.279(07)$ &  &  &
 $2.702(-05)$ & $1.645(-04)$& & \\
$4-19$ & $1018.526$ &  $8.703(06)$ & $2.181(06)$ & $2.75(06)$ &  &
 $5.415(-03)$ & $1.326(-03)$&$1.336(-03)$ & \\
$5-19$ & $1063.250$ &  $9.488(07)$ & $2.892(06)$ &  &  &
 $6.433(-02)$ & $1.952(-03)$&$1.862(-03)$ & \\
$6-19$ & $1063.250$ &  $1.100(05)$ & $1.283(06)$ & $1.22(06)$ &  &
 $7.545(-05)$ & $8.336(-04)$& $7.961(-04)$&  \\
$7-19$ & $1228.189$ & $9.754(07)$ & $4.377(08)$ & $4.54(08)$ &  &
 $8.824(-02)$ & $3.701(-01)$&$3.845(-01)$ & $3.66(-01)$ \\
$8-19$ & $1233.748$ &  $6.973(08)$ & $3.698(08)$ & $3.48(08)$ &  &
 $6.365(-01)$ & $3.420(-01)$& $3.221(-01)$& \\
$9-19$ & $2251.333$ &  $3.457(05)$ & $6.305(06)$ & $7.30(06)$ &  &
 $1.051(-03)$ & $1.648(-02)$&$1.887(-02)$ & \\
$10-19$ & $2271.969$ & $4.466(04)$ & $2.421(06)$ & $2.63(06)$ &  &
 $1.382(-04)$ & $6.344(-03)$&$6.807(-03)$ & \\
$3-20$ & $148.267$ &  $1.378(08)$ & $8.949(07)$ &  &  &
 $9.085(-04)$ & $5.760(-04)$& &  \\
$5-20$ & $1058.626$ & $6.333(05)$ & $1.717(05)$ &  &  &
 $2.128(-04)$ & $5.804(-05)$& & \\
$6-20$ & $1064.537$ & $6.918(06)$ & $1.331(06)$ &  &  &
 $2.351(-03)$ & $4.332(-04)$& &  \\
$7-20$ & $1222.023$ & $5.861(08)$ & $6.643(08)$ &  &  &
 $2.624(-01)$ & $2.813(-01)$& &  \\
$8-20$ & $1227.526$ &  $6.666(07)$ & $5.304(07)$ &  &  &
 $3.012(-02)$ & $2.458(-02)$& &  \\
$10-20$ & $2250.959$ & $2.878(07)$ & $2.829(07)$ &  &  &
 $4.373(-02)$ & $3.719(-02)$& &  \\
$3-21$ & $147.538$ & $1.880(08)$ & $1.114(08)$ & $1.10(08)$ &
& $2.454(-03)$ & $1.422(-03)$&$1.339(-03)$ & \\
$4-21$ & $981.153$ &  $1.120(07)$ & $7.761(06)$ & $5.01(06)$ &  &
 $6.464(-03)$ & $4.476(-03)$&$2.904(-03)$ &  \\
$5-21$ & $1022.589$ & $4.986(06)$ & $1.343(07)$ & $1.08(07)$ &  &
 $3.127(-03)$ & $8.554(-03)$& $6.902(-03)$& \\
$6-21$ & $1028.103$ &  $1.619(07)$ & $2.197(06)$ &  &  &
 $1.026(-02)$ & $1.348(-03)$& &  \\
$7-21$ & $1174.254$ &  $5.806(08)$ & $2.535(08)$ & $2.33(08)$ &  &
 $4.801(-01)$ & $2.008(-01)$&$1.853(-01)$ & $2.24(-01)$ \\
$8-21$ & $1179.334$ & $4.876(07)$ & $4.094(08)$ & $4.31(08)$ &  &
 $4.067(-02)$ & $3.538(-01)$& $3.741(-01)$& $3.42(-01)$ \\
$9-21$ & $2076.504$ &  $4.517(07)$ & $4.361(07)$ & $4.31(07)$ &  &
 $1.168(-01)$ & $1.018(-01)$& $1.002(-01)$& \\
$10-21$ & $2094.047$ & $5.294(06)$ & $3.415(06)$ & $3.31(06)$ &  &
 $1.392(-02)$ & $7.996(-03)$& $7.726(-03)$&  \\
$1-22$ & $92.226$ & $2.083(10)$ & $1.964(10)$ & $1.96(10)$ &
& $5.314(-02)$ & $4.908(-02)$&$4.920(-02)$ & $6.70(-02)$ \\
$2-22$ & $92.468$ &  $1.110(10)$ & $1.232(10)$ & $1.23(10)$ &
&  $2.845(-02)$ & $3.108(-02)$&$3.097(-02)$ & $4.26(-02)$ \\
$3-23$ & $146.805$ &  $5.237(06)$ & $1.548(05)$ &  &  &
 $6.769(-05)$ & $1.968(-06)$& & \\
$4-23$ & $949.626$ &  $7.718(08)$ & $6.774(08)$ & $6.81(08)$ &  &
 $4.174(-01)$ & $3.808(-01)$& $3.837(-01)$& $3.48(-01)$ \\
$5-23$ & $988.389$ & $3.575(08)$ & $3.063(08)$ & $3.08(08)$ &  &
 $2.095(-01)$ & $1.899(-01)$&$1.909(-01)$ & $1.89(-01)$ \\
$6-23$ & $993.539$ &  $5.751(08)$ & $5.521(08)$ & $5.52(08)$ &  &
 $3.405(-01)$ & $3.301(-01)$&$3.303(-01)$ & $3.30(-01)$ \\
$7-23$ & $1129.379$ & $4.435(06)$ & $4.529(05)$ & $5.45(04)$ &  &
 $3.393(-03)$ & $3.484(-04)$&$4.197(-05)$ & \\
$8-23$ & $1134.078$ &  $3.772(06)$ & $7.482(06)$ & $5.32(06)$ &  &
 $2.909(-03)$ & $6.270(-03)$&$4.456(-03)$ &  \\
$9-23$ & $1940.178$ & $1.976(06)$ & $1.302(06)$ &  &  &
 $4.461(-03)$ & $2.892(-03)$& &  \\
$10-23$ & $1955.485$ & $2.723(04)$ & $9.887(04)$ &  &  &
 $6.245(-05)$ & $2.200(-04)$& &  \\
$3-24$ & $146.615$ & $4.880(07)$ & $2.590(07)$ &  &  &
 $3.145(-04)$ & $1.644(-04)$& & \\
$5-24$ & $979.803$ & $4.553(07)$ & $2.065(06)$ &  &  &
 $1.311(-02)$ & $6.328(-04)$& & \\
$6-24$ & $984.864$ & $1.963(03)$ & $3.411(06)$ &  &  &
 $5.710(-07)$ & $1.008(-03)$& &  \\
$7-24$ & $1118.182$ &  $2.447(08)$ & $1.833(08)$ &  &  &
 $9.173(-02)$ & $6.964(-02)$& &  \\
$8-24$ & $1122.788$ & $7.678(08)$ & $6.707(08)$ &  &  &
 $2.902(-01)$ & $2.772(-01)$& & \\
$10-24$ & $1922.160$ & $1.678(07)$ & $2.055(07)$ &  &  &
 $1.859(-02)$ & $2.237(-02)$& &  \\
$5-25$ & $735.260$ &  $4.657(06)$ &  &  &  &
$2.265(-03)$ & & & \\
$6-25$ & $761.314$ & $5.428(05)$ & $1.357(04)$ &$1.51(04)$&  &
 $2.830(-04)$ & $7.164(-06)$&$8.035(-06)$ &  \\
$7-25$ & $838.603$ &  $7.411(04)$ & $7.983(04)$ &  &  &
$4.688(-05)$ & $5.100(-05)$& & \\
$9-25$ & $1215.905$ & $8.749(07)$ & $7.805(07)$ & $7.65(07)$ &  &
 $1.163(-01)$ & $1.035(-01)$& $1.020(-01)$& $1.09(-01)$ \\
$10-25$ & $1221.899$ &  $7.763(08)$ & $7.612(08)$ & $7.56(08)$ &
&$1.043(+00)$ & $1.011(+00)$&1.006(+00) & $1.11(+00)$ \\
$4-26$ & $733.553$ & $8.310(05)$ & $1.120(05)$ &$1.20(05)$  &  &
 $5.363(-04)$ & $7.446(-05)$&$8.035(-05)$ & \\
$9-26$ & $1211.242$ & $8.876(08)$ & $8.628(08)$ & $8.56(08)$ &  &
 $1.562(+00)$ & $1.501(+00)$&$1.499(+00)$ & $1.64(+00)$ \\
$3-27$ & $138.443$ & $1.456(03)$ & $4.078(06)$ & $4.89(06)$ &
& $1.674(-08)$ & $4.622(-05)$& & \\
$4-27$ & $682.827$ & $3.924(05)$ & $6.054(04)$ &$6.24(04)$  &  &
 $1.097(-04)$ & $1.749(-05)$&$1.811(-05)$ & \\
$5-27$ & $702.641$ & $1.011(06)$ & 5.116(-1) &$1.89(00)$  &  &
 $2.994(-04)$ & $1.584(-10)$&$5.902(-10)$ & \\
$6-27$ & $705.240$ & $1.533(06)$ & $1.283(06)$ & $4.40(05)$ &  &
 $4.573(-04)$ & $1.294(-04)$&$1.339(-04)$ &  \\
$7-27$ & $771.072$ &  $1.238(06)$ & $1.967(07)$ &$2.00(07)$  &  &
 $4.413(-04)$ & $7.076(-03)$&$7.244(-03)$ & \\
$8-27$ & $773.259$ & $5.680(06)$ & $5.306(06)$ &$5.19(06)$  &  &
 $2.037(-03)$ & $2.024(-03)$&$1.995(-03)$ & \\
$9-27$ & $1078.900$ &  $1.254(08)$ & $7.247(07)$ & $7.17(07)$ &  &
 $8.756(-02)$ & $5.049(-02)$& $5.023(-02)$& \\
$10-27$ & $1083.616$ &  $1.118(09)$ & $1.117(09)$ & $1.11(09)$ &
& $7.875(-01)$ & $7.792(-01)$& $7.798(-01)$& $8.03(-01)$ \\
$4-28$ & $679.840$ & $2.305(07)$ & $4.734(05)$ &$4.88(05)$  &  &
 $9.584(-03)$ & $2.040(-04)$&$2.118(-04)$ & \\ \hline
\end{tabular}
\end{center}
\end{table*}
\newpage
\addtocounter{table}{-1}
\begin{table*}
\caption{$Continued.$}
\label{tab:2}       
\begin{center}
\begin{tabular}{lrllllllll}
\hline
$Transition$ & $\lambda ({%
{\AA}%
})$   &$A_{ki}(s^{-1})$  &  &  &  &$gf$  &  \\
\cline{3-5}\cline{7-10} &  &  $present$ & $FF$ & $NIST$ & &
$present$ &$FF$ &$NIST$ & $CT$ \\ \hline $6-28$ & $702.055$ &
$4.910(06)$ & $5.481(03)$ &$4.62(03)$ & &
 $2.177(-03)$ & $2.467(-06)$&$2.094(-06)$ &  \\
$7-28$ & $767.266$ & $6.267(06)$ & $1.070(07)$ &$1.05(07)$  &  &
 $3.319(-03)$ & $5.740(-03)$&$5.675(-03)$& \\
$9-28$ & $1071.462$ &  $1.150(09)$ & $1.100(09)$ & $1.10(09)$ &  &
 $1.187(+00)$ & $1.140(+00)$& $1.140(+00)$& $1.17(+00)$ \\
$10-28$ & $1076.114$ & $1.288(08)$ & $1.137(08)$ & $1.13(08)$ &  &
 $1.342(-01)$ & $1.179(-01)$&$1.174(-01)$ & $1.18(-01)$ \\
$3-29$ & $135.238$ &  $5.172(08)$ & $7.112(08)$ & $7.11(08)$&
& $5.673(-03)$ & $7.758(-03)$&$7.762(-03)$ &  \\
$4-29$ & $611.374$ &  $2.805(07)$ & $5.525(05)$ &$5.48(05)$  &  &
 $6.287(-03)$ & $1.326(-04)$&$1.327(-04)$ & \\
$5-29$ & $627.211$ &  $2.072(07)$ & $1.148(05)$ &$1.09(05)$  &  &
 $4.888(-03)$ & $2.936(-05)$&$2.818(-05)$ &  \\
$6-29$ & $629.281$ & $2.656(05)$ & $1.930(07)$ &$1.92(07)$ &  &
$6.306(-05)$ & $4.820(-03)$&$4.841(-03)$ &  \\
$7-29$ & $681.173$ & $1.678(09)$ & $1.402(09)$ &$1.38(09)$  &  &
 $4.670(-01)$ & $4.104(-01)$&$4.083(-01)$ & $4.32(-01)$ \\
$8-29$ & $682.880$ &  $3.009(08)$ & $2.957(08)$ &$2.92(08)$  &  &
 $8.415(-02)$ & $9.122(-02)$&$9.078(-02)$ & $9.61(-02)$ \\
$9-29$ & $910.722$ &  $1.294(09)$ & $1.235(09)$ & $1.23(09)$ &  &
 $6.436(-01)$ & $6.498(-01)$&$6.486(-01)$& \\
$10-29$ & $914.081$ & $1.412(08)$ & $5.801(07)$ & $5.69(07)$ &  &
 $7.075(-02)$ & $3.054(-02)$&$3.013(-02)$ & \\
$3-30$ & $135.230$ &  $5.478(08)$ & $8.058(08)$ & $8.06(08)$ &
 &$3.004(-03)$ & $4.368(-03)$&$4.355(-03)$ &  \\
$5-30$ & $627.022$ & $7.195(07)$ & $3.778(06)$ &$3.77(06)$  &  &
 $8.482(-03)$ & $4.686(-04)$&$4.720(-04)$ &  \\
$6-30$ & $629.090$ & $1.571(05)$ & $4.630(06)$ &$4.61(06)$  &  &
 $1.864(-05)$ & $5.608(-04)$&$5.636(-04)$ &  \\
$7-30$ & $680.950$ & $6.378(08)$ & $4.582(08)$ &$4.52(08)$ &  &
 $8.868(-02)$ & $6.496(-02)$&$6.411(-02)$ & $6.83(-02)$ \\
$8-30$ & $682.655$ & $1.200(09)$ & $1.070(09)$ &$1.06(09)$&  &
 $1.677(-01)$ & $1.596(-01)$&$1.588(-01)$ & $1.68(-01)$ \\
$10-30$ & $913.679$ &  $1.459(09)$ & $1.448(09)$ & $1.44(09)$ &  &
 $3.653(-01)$ & $3.649(-01)$&$3.647(-01)$ & $3.73(-01)$ \\
$3-31$ & $130.708$ &  $8.779(08)$ & $6.247(08)$ &  &  &
 $8.995(-03)$ & $6.224(-03)$& & \\
$4-31$ & $528.555$ &  $4.887(06)$ & $1.410(05)$ &  &  &
 $8.188(-04)$ & $2.283(-05)$& &  \\
$5-31$ & $540.350$ &  $3.953(06)$ & $1.024(04)$ &  &  &
 $6.922(-04)$ & $1.745(-06)$& &  \\
$6-31$ & $541.886$ & $6.974(06)$ & $2.069(06)$ &  &  &
 $1.228(-03)$ & $3.461(-04)$& & \\
$7-31$ & $579.929$ & $2.418(08)$ & $1.207(08)$ &  &  &
 $4.877(-02)$ & $2.297(-02)$& &  \\
$8-31$ & $581.166$ & $6.960(07)$ & $2.686(07)$ &  &  &
 $1.410(-02)$ & $5.334(-03)$& &  \\
$9-31$ & $738.377$ & $1.559(05)$ & $2.646(06)$ &  &  &
 $5.098(-05)$ & $7.962(-04)$& &  \\
$10-31$ & $740.583$ & $2.238(04)$ & $1.572(06)$ &  &  &
 $7.363(-06)$ & $4.732(-04)$& &  \\
$3-32$ & $130.562$ & $8.400(08)$ & $5.510(08)$ &  &  &
 $4.293(-03)$ & $2.744(-03)$& &  \\
$5-32$ & $537.865$ & $8.805(06)$ & $6.903(05)$ &  &  &
 $7.638(-04)$ & $5.880(-05)$& & \\
$6-32$ & $539.386$ &  $1.317(03)$ & $7.534(05)$ &  &  &
 $1.149(-07)$ & $6.296(-05)$& &  \\
$7-32$ & $577.068$ &  $8.320(07)$ & $7.216(07)$ &  &  &
 $8.308(-03)$ & $6.860(-03)$& &  \\
$8-32$ & $578.292$ &  $1.931(08)$ & $1.451(08)$ &  &  &
 $1.937(-02)$ & $1.438(-02)$& &  \\
$10-32$ & $735.923$ & $5.602(06)$ & $1.894(06)$ &  &  &
 $9.097(-04)$ & $2.847(-04)$& & \\
$1-34$ & $85.232$ & $8.382(06)$ & $2.275(07)$ & $2.30(07)$ &
&  $5.477(-05)$ & $1.464(-04)$&$1.482(-04)$ & \\
$1-35$ & $85.224$ & $9.086(05)$ & $1.456(08)$ & $1.45(08)$ &
& $3.958(-06)$ & $6.244(-04)$&$6.237(-04)$ &  \\
$2-35$ & $85.431$ & $6.482(07)$ & $3.464(07)$ & $3.50(07)$ &
&  $2.837(-04)$ & $1.498(-04)$&$1.517(-04)$ &  \\
$1-36$ & $85.181$ & $8.300(07)$ & $8.691(07)$ & $8.61(07)$ &
& $1.806(-04)$ & $1.861(-04)$&$1.845(-04)$ & \\
$2-36$ & $85.387$ & $1.621(08)$ & $2.003(08)$ & $1.99(08)$ &
& $3.545(-04)$ & $4.328(-04)$&$4.295(-04)$ &  \\
$1-38$ & $84.422$ & $8.666(07)$ & $1.172(09)$ & $1.18(09)$ &
& $3.704(-04)$ & $4.921(-03)$&$4.954(-03)$ & $1.03(-01)$ \\
$2-38$ & $84.625$ &$4.078(08)$ & $2.408(09)$ & $2.44(09)$ &
& $1.751(-03)$ & $1.020(-02)$&$1.035(-02)$ & $1.96(-02)$ \\
$1-41$ & $84.133$ &  $8.295(08)$ & $1.409(09)$ &  &
& $1.761(-03)$ & $2.958(-03)$& &  \\
$2-41$ & $84.334$ & $5.368(08)$ & $1.743(07)$ &  &
& $1.145(-03)$ & $3.690(-05)$& & \\
$1-42$ & $84.066$ & $4.006(08)$ & $2.740(09)$ &  &
& $1.698(-03)$ & $1.148(-02)$& &  \\
$2-42$ & $84.267$ & $3.639(07)$ & $5.164(08)$ &  &
& $1.550(-04)$ & $2.182(-03)$& & $3.00(-04)$ \\
$1-43$ & $83.927$ & $1.359(08)$ & $3.789(08)$ &  &  &
 $8.612(-04)$ & $2.374(-03)$& & $1.46(-02)$ \\
$1-45$ & $83.974$ &$1.817(10)$ & $2.211(10)$ &  &
&$1.153(-01)$ & $1.386(-01)$& & \\
$1-46$ & $83.742$ & $3.143(10)$ & $2.980(10)$ & $2.99(10)$ &
&$6.608(-02)$ & $6.196(-02)$&$6.194(-02)$ & $5.95(-02)$ \\
$2-46$ & $83.942$ & $5.476(10)$ & $4.368(10)$ & $4.36(10)$ &
& $1.157(-01)$ &$9.156(-02)$ &$9.141(-02)$ & $8.40(-02)$ \\
$1-47$ & $83.546$ & $6.063(10)$ & $1.686(10)$ & $1.77(10)$ &
& $2.538(-01)$ & $6.960(-02)$&7.227(-02) & $3.09(-01)$ \\
$2-47$ & $83.745$ & $2.522(10)$ & $9.938(10)$ & $9.74(10)$ &
&$1.061(-01)$ & $4.138(-01)$&$4.064(-01)$ &$4.00(-04)$ \\
$1-48$ & $83.678$ & $4.790(10)$ & $9.699(10)$ & $9.60(10)$ &
& $2.012(-01)$ & $4.028(-01)$&$3.990(-01)$ & $7.03(-02)$ \\
$2-48$ & $83.877$ & $1.934(11)$ & $9.296(10)$ & $9.47(10)$ &
& $8.162(-01)$ & $3.894(-01)$&$3.981(-01)$ & $8.26(-01)$ \\
$1-49$ & $83.595$ & $2.281(11)$ & $2.341(11)$ & $2.34(11)$ &
& $1.434(+00)$ &  $1.545(+00)$&$1.452(+00)$&$1.27(+00)$ \\
$1-52$ & $81.031$ & $3.016(11)$ & $2.850(11)$ & $2.76(11)$ &
& $5.939(-01)$ & $5.556(-01)$&$5.395(-01)$ &$6.00(-01)$ \\
$2-52$ & $81.218$ &$7.742(10)$ & $7.643(10)$ & $8.45(10)$ &
& $1.531(-01)$ & $1.502(-01)$&$1.663(-01)$ & $2.13(-01)$ \\
$1-53$ & $80.984$ &$3.209(11)$ & $2.749(11)$ & $2.74(11)$ &
& $1.262(+00)$ & $1.068(+00)$&$1.069(+00)$ & $1.33(+00)$ \\
$2-53$ & $81.171$ & $3.233(10)$ & $5.658(10)$ & $5.67(10)$ &
& $1.278(-01)$ & $2.218(-01)$&$2.218(-01)$ & $2.91(-01)$ \\
$1-54$ & $80.914$ & $6.868(10)$ & $6.732(10)$ & $6.85(10)$ &
& $1.348(-01)$ & $1.306(-01)$&$1.333(-01)$ & $2.11(-01)$ \\
$2-54$ & $81.100$ &$2.860(11)$ & $2.804(11)$ & $2.78(11)$ & &
$5.640(-01)$ & $5.484(-01)$&$5.457(-01)$ & $6.27(-01)$ \\ \hline
\end{tabular}
\end{center}
\end{table*}
\newpage
\addtocounter{table}{-1}
\begin{table*}
\caption{$Continued.$}
\label{tab:2}       
\begin{center}
\begin{tabular}{lrllllllll}
\hline
$Transition$ & $\lambda ({%
{\AA}%
})$  &$A_{ki}(s^{-1})$   &  &  &  &$gf$  &  \\
\cline{3-5}\cline{7-10} &   & $present$ & $FF$ & $NIST$ &  &
$present$ & $FF$ &$NIST$& $CT$ \\ \hline $1-56$ & $80.844$ &
$3.445(10)$ & $6.766(10)$ &  &
& $2.025(-01)$ & $3.939(-01)$& & $7.83(-02)$ \\
$1-57$ & $80.800$ &  $1.848(11)$ & $1.335(11)$ & $1.31(11)$ &
& $1.085(+00)$ & $7.764(-01)$&$7.603(-01)$ & $1.59(+00)$ \\
$1-58$ & $80.755$ & $1.352(10)$ & $3.492(10)$ & $3.50(10)$ &
& $5.289(-02)$ & $1.351(-01)$&$1.355(-01)$ & $1.97(-01)$ \\
$2-58$ & $80.941$ & $1.969(11)$ & $1.891(11)$ & $1.89(11)$ &
& $7.735(-01)$ & $7.380(-01)$&$7.362(-01)$ & $1.01(+00)$ \\
$1-59$ & $78.069$ & $8.919(09)$ & $7.731(09)$ & $7.74(09)$ &
& $3.260(-02)$ & $2.776(-02)$&$2.779(-02)$ &  \\
$2-59$ & $78.242$ & $4.642(10)$ & $5.577(10)$ & $5.57(10)$ &
& $1.704(-01)$ & $2.018(-01)$&$2.018(-01)$ & $2.72(-01)$ \\
$1-60$ & $78.025$ &  $4.737(10)$ & $5.054(10)$ & $5.05(10)$ &
& $2.594(-01)$ & $2.723(-01)$&$2.722(-01)$ & $3.75(-01)$ \\
$1-61$ & $76.060$ &  $2.957(07)$ &  &  &  &
$1.539(-04)$ & & &  \\
$1-62$ & $75.899$ & $3.236(09)$ &  &  &  &
$5.589(-03)$ & & & \\
$2-62$ & $76.063$ &  $6.405(09)$ &  &  &  &
$1.111(-02)$ & & & \\
$1-63$ & $75.797$ & $1.243(09)$ &  &  &  &
$4.281(-03)$ & & & $1.14(-02)$ \\
$2-63$ & $75.961$ &  $2.661(08)$ &  &  &  &$9.209(-04)$ & & &  \\
$1-64$ & $75.676$ & $2.243(10)$ &  & $1.24(10)$ &  &
 $7.702(-02)$ & &$4.197(-02)$ & $4.21(-02)$ \\
$2-64$ & $75.839$ & $4.741(09)$ &  & $2.11(09)$ &  & $1.635(-02)$ & &$7.194(-03)$ &$7.20(-03)$ \\
$1-65$ & $75.583$ & $5.689(09)$ &  &  &  &  $%
9.746(-03)$ & & & \\
$2-65$ & $75.746$ &  $1.202(10)$ &  &  &  &  $%
2.069(-02)$ & & & \\
$4-66$ & $278.937$ & $1.029(10)$ &  &  &  &  $%
7.204(-01)$ & & &  \\
$6-66$ & $282.606$ &  $2.931(09)$ &  &  &  &  $%
2.106(-01)$ & & &  \\
$7-66$ & $292.617$ &  $2.511(07)$ &  &  &  &  $%
1.934(-03)$ & & &  \\
$9-66$ & $328.147$ &  $5.179(07)$ &  &  &  &  $%
5.016(-03)$ & & & \\
$10-66$ & $328.582$ &  $6.473(03)$ &  &  &  &  $%
6.287(-07)$ & & & \\
$3-67$ & $107.011$ &  $3.996(06)$ &  &  &  &  $%
2.744(-05)$ & & &  \\
$4-67$ & $278.853$ &  $6.399(09)$ &  &  &  &  $%
2.984(-01)$ & & &  \\
$5-67$ & $282.101$ &  $3.730(09)$ &  &  &  &  $%
1.780(-01)$ & & & \\
$6-67$ & $282.519$ &  $1.629(09)$ &  &  &  &  $%
7.796(-02)$ & & & \\
$7-67$ & $292.524$ & $1.005(07)$ &  &  &  &  $%
5.159(-04)$ & & & \\
$8-67$ & $292.839$ &  $4.215(08)$ &  &  &  &  $%
2.168(-02)$ & & &  \\
$9-67$ & $328.031$ &  $2.309(07)$ &  &  &  & $%
1.490(-03)$ & & & \\
$10-67$ & $328.466$ &  $2.775(06)$ &  &  &  &  $%
1.796(-04)$ & & &  \\
$3-68$ & $106.811$ &  $3.368(06)$ &  &  &  &  $%
1.152(-05)$ & & & \\
$5-68$ & $280.715$ & $1.075(09)$ &  &  &  &  $%
2.539(-02)$ &  & & \\
$6-68$ & $281.129$ & $1.059(10)$ &  &  &  &  $%
2.511(-01)$ & & & \\
$7-68$ & $291.034$ & $3.005(07)$ &  &  &  & $%
7.632(-04)$ & & &  \\
$8-68$ & $291.345$ & $6.976(07)$ &  &  &  &  $%
1.776(-03)$ & & & \\
$10-68$ & $326.587$ & $8.116(06)$ &  &  &  &  $%
2.596(-04)$ & & &  \\
$4-69$ & $273.806$ &  $3.264(09)$ &  &  &  &  $%
2.935(-01)$ & & &  \\
$9-69$ & $321.069$ &  $7.827(06)$ &  &  &  &  $%
9.678(-04)$ & & &  \\
$3-70$ & $105.978$ & $6.730(03)$ &  &  &  &  $%
2.266(-08)$ & & & \\
$5-70$ & $275.031$ & $3.121(09)$ &  &  &  &  $%
7.079(-02)$ & & &  \\
$6-70$ & $275.428$ &  $3.863(02)$ &  &  &  & $%
8.788(-09)$ & & &  \\
$7-70$ & $284.929$ & $8.093(04)$ &  &  &  &  $%
1.970(-06)$ & & & \\
$8-70$ & $285.227$ &  $1.165(08)$ &  &  &  & $%
2.841(-03)$ & & &  \\
$10-70$ & $318.919$ &  $1.142(04)$ &  &  &  & $%
3.482(-07)$ & & &  \\
$3-71$ & $105.843$ & $1.511(06)$ &  &  &  & $%
1.011(-05)$ & & & \\
$4-71$ & $271.056$ & $9.192(06)$ &  &  &  & $%
4.050(-04)$ & & & \\
$5-71$ & $274.125$ & $1.417(09)$ &  &  &  & $%
6.385(-02)$ & & &  \\
$6-71$ & $274.519$ & $1.151(09)$ &  &  &  & $%
5.201(-02)$ & & & \\
$7-71$ & $283.956$ & $3.861(04)$ &  &  &  & $%
1.867(-06)$ & & & \\
$8-71$ & $284.252$ & $2.438(08)$ &  &  &  & $%
1.181(-02)$ & & &  \\
$9-71$ & $317.295$ & $4.312(06)$ &  &  &  & $%
2.603(-04)$ & & &  \\
$10-71$ & $317.701$ & $6.615(04)$ &  &  &  & $%
4.004(-06)$ & & & \\
$4-72$ & $270.433$ & $5.068(07)$ &  &  &  & $%
3.334(-03)$ & & & \\
$6-72$ & $273.880$ & $1.299(09)$ &  &  &  & $%
8.765(-02)$ & & &  \\
$7-72$ & $283.272$ & $2.124(09)$ &  &  &  & $%
1.533(-01)$ & & &  \\
$9-72$ & $316.441$ & $3.720(04)$ &  &  &  & $%
3.351(-06)$ & & &  \\
$10-72$ & $316.845$ & $3.494(05)$ &  &  &  & $%
3.155(-05)$ & & & \\
$4-73$ & $271.104$ & $2.072(08)$ &  &  &  & $%
1.370(-02)$ & & & \\
$6-73$ & $274.569$ & $1.876(09)$ &  &  &  & $%
1.272(-01)$ & & & \\ \hline
\end{tabular}
\end{center}
\end{table*}
\newpage
\addtocounter{table}{-1}
\begin{table*}
\caption{$Continued.$}
\label{tab:2}       
\begin{center}
\begin{tabular}{lrllllllll}
\hline
$Transition$ & $\lambda ({%
{\AA}%
})$ &$A_{ki}(s^{-1})$  &  &  &  & $gf$  &  \\
\cline{3-5}\cline{7-10} &  & $present$ & $FF$ & $NIST$ &  &
$present$ &$FF$ &$NIST$ & $CT$ \\ \hline
$7-73$ & $284.009$ & $8.893(08)$ &  &  &  & $%
6.453(-02)$ & & &  \\
$9-73$ & $317.361$ & $2.303(06)$ &  &  &  & $%
2.086(-04)$ & & &  \\
$10-73$ & $317.768$ & $6.699(06)$ &  &  &  & $%
6.085(-04)$ & & &  \\
$3-74$ & $105.742$ & $2.059(05)$ &  &  &  & $%
1.381(-06)$ & & &  \\
$4-74$ & $270.396$ & $1.558(08)$ &  &  &  & $%
6.832(-03)$ & & &  \\
$5-74$ & $273.449$ & $5.931(08)$ &  &  &  & $%
2.660(-02)$ & & &  \\
$6-74$ & $273.842$ & $6.104(05)$ &  &  &  & $%
2.745(-05)$ & & & \\
$7-74$ & $283.232$ & $3.949(08)$ &  &  &  & $%
1.900(-02)$ & & &  \\
$8-74$ & $283.526$ & $2.129(09)$ &  &  &  & $%
1.026(-01)$ & & &  \\
$9-74$ & $316.390$ & $1.947(06)$ &  &  &  & $%
1.169(-04)$ & & &  \\
$10-74$ & $316.795$ & $1.613(06)$ &  &  &  & $%
9.709(-05)$ & & & \\
$3-75$ & $105.683$ & $1.462(07)$ &  &  &  & $%
4.896(-05)$ & & & \\
$5-75$ & $273.056$ & $1.449(06)$ &  &  &  & $%
3.239(-05)$ & & &  \\
$6-75$ & $273.447$ & $1.060(07)$ &  &  &  & $%
2.376(-04)$ & & &  \\
$7-75$ & $282.809$ & $2.598(09)$ &  &  &  & $%
6.230(-02)$ & & &  \\
$8-75$ & $283.103$ & $5.264(07)$ &  &  &  & $%
1.265(-03)$ & & & \\
$10-75$ & $316.266$ & $4.386(08)$ &  &  &  & $%
1.315(-02)$ & & &  \\
$3-76$ & $105.361$ & $2.021(06)$ &  &  &  & $%
1.345(-05)$ & & & \\
$4-76$ & $267.917$ & $3.085(08)$ &  &  &  & $%
1.328(-02)$ & & &  \\
$5-76$ & $270.915$ & $1.073(09)$ &  &  &  & $%
4.722(-02)$ & & &  \\
$6-76$ & $271.300$ & $1.127(09)$ &  &  &  & $%
4.973(-02)$ & & &  \\
$7-76$ & $280.513$ & $1.603(07)$ &  &  &  & $%
7.565(-04)$ & & &  \\
$8-76$ & $280.802$ & $3.923(07)$ &  &  &  & $%
1.855(-03)$ & & & \\
$9-76$ & $313.002$ & $1.264(06)$ &  &  &  & $%
7.425(-05)$ & & & \\
$10-76$ & $313.398$ & $1.091(06)$ &  &  &  & $%
6.425(-05)$ & & &  \\
$3-77$ & $104.974$ & $4.741(07)$ &  &  &  & $%
3.133(-04)$ & & &  \\
$4-77$ & $265.429$ & $3.542(06)$ &  &  &  & $%
1.497(-04)$ & & &  \\
$5-77$ & $268.371$ & $6.593(07)$ &  &  &  & $%
2.848(-03)$ & & & \\
$6-77$ & $268.750$ & $7.602(06)$ &  &  &  & $%
3.293(-04)$ & & &  \\
$7-77$ & $277.787$ & $1.359(09)$ &  &  &  & $%
6.288(-02)$ & & & \\
$8-77$ & $278.07$ & $1.709(08)$ &  &  &  & $%
7.925(-03)$ & & & \\
$9-77$ & $309.612$ & $1.217(09)$ &  &  &  & $%
6.997(-02)$ & & & \\
$10-77$ & $309.999$ & $1.274(08)$ &  &  &  & $%
7.344(-03)$ & & &  \\
$3-78$ & $104.935$ & $3.845(07)$ &  &  &  & $%
1.270(-04)$ & & &  \\
$5-78$ & $268.119$ & $1.251(08)$ &  &  &  & $%
2.696(-03)$ & & & \\
$6-78$ & $268.497$ & $1.917(07)$ &  &  &  & $%
4.144(-04)$ & & &  \\
$7-78$ & $277.517$ & $1.251(06)$ &  &  &  & $%
2.888(-05)$ & & & \\
$8-78$ & $277.800$ & $1.619(09)$ &  &  &  & $%
3.747(-02)$ & & &  \\
$10-78$ & $309.663$ & $8.910(08)$ &  &  &  & $%
2.562(-02)$ & & & \\
$1-79$ & $73.253$ & $9.110(09)$ &  & $8.24(09)$ &  &  $4.398(-02)$ & &$3.935(-02)$ & $3.94(-02)$ \\
$1-80$ & $73.232$ & $1.305(09)$ &  &  &  &  $%
4.198(-03)$ & & &  \\
$2-80$ & $73.385$ &  $7.299(09)$ &  &  &  &  $%
2.357(-02)$ & & &  \\
$1-82$ & $72.246$ & $2.920(08)$ &  &  &  & $%
1.371(-03)$ & & &  \\
$1-83$ & $72.222$ & $1.216(02)$ &  &  &  &  $%
3.802(-10)$ & & & \\
$2-83$ & $72.371$ & $2.423(08)$ &  &  &  & $%
7.611(-04)$ & & &  \\
$1-84$ & $72.175$ & $5.005(08)$ &  &  &  & $%
7.818(-04)$ & & & \\
$2-84$ & $72.323$ & $9.647(08)$ &  &  &  & $%
1.513(-03)$ & & &  \\
$1-87$ & $71.976$ & $2.262(10)$ &  &  &  & $%
1.054(-01)$& &  & $1.89(-01)$ \\
$1-88$ & $71.976$ & $1.382(08)$ &  &  &  & $%
4.293(-04)$ & & &  \\
$2-88$ & $72.124$ & $6.586(08)$ &  &  &  & $%
2.055(-03)$ & & & $1.03(-01)$ \\
$1-89$ & $71.910$ & $2.267(09)$ &  &  &  & $%
3.515(-03)$ & & &  \\
$2-89$ & $72.057$ & $3.055(09)$ &  &  &  & $%
4.756(-03)$ & & &  \\
$1-90$ & $71.845$ & $1.208(09)$ &  &  &  & $%
3.740(-03)$ & & & $2.35(-02)$ \\
$2-90$ & $71.991$ & $2.033(07)$ &  &  &  & $%
6.318(-05)$ & & & $2.00(-04)$ \\
$1-91$ & $71.742$ & $1.462(09)$ &  &  &  & $%
6.769(-03)$ & & & $1.96(-02)$ \\
$1-92$ & $71.793$ & $1.906(10)$ &  &  &  & $%
8.839(-02)$ & & & \\
$1-94$ & $71.761$ & $2.760(10)$ &  & $2.19(10)$ &  &
$4.261(-02)$ & &$3.341(-02)$ & $3.34(-02)$ \\
$2-94$ & $71.907$ & $5.142(10)$ &  & $3.56(10)$ &  & $7.972(-02)$&
& $5.457(-02)$& $5.47(-02)$ \\
$1-95$ & $71.761$ & $2.760(10)$ &  & $7.20(10)$ &  & $4.261(-02)$&
& & \\
$2-95$ & $71.907$ & $5.142(10)$ &  & $3.26(07)$ &  & $7.972(-02)$&
&$1.000(-04)$ &$1.00(-04)$\\\hline
\end{tabular}
\end{center}
\end{table*}

\begin{table*}
\caption{This table gives electron-, proton and singly-charged
helium-impact broadening parameters for Si VI lines calculated
using SUPERSTRUCTURE oscillator strength, for a perturber density
of 10$^{17}$ cm$^{-3}$ and temperature of 50000 to 800000 K.
Transitions, averaged wavelength for the multiplet (in \AA) and
parameter \emph{C} are also given. This parameter when divided
with the corresponding Stark width gives an estimate for the
maximal pertuber density for which the line may be treated as
isolated. $w_{e}$: electron-impact full Stark width at half
maximum, $d_{e}$: electron-impact Stark shift, $w_{H^{+}}$:
proton-impact full Stark width at half maximum, $d_{H^{+}}$:
proton-impact Stark shift, $w_{He^{+}}$: singly charged
helium-impact full Stark width at half maximum, $d_{He^{+}}$:
singly charged helium-impact Stark shift. $w_{MSE}$:
electron-impact full Stark width at half maximum calculated by
\citet{b26} using modified semiempirical formula \citep{b7}.}
\label{tab:2}       
\begin{center}
\begin{tabular}{crrrrrrrr}
\hline transition  & T(K) & $w_{e}$ & $d_{e}$ & $w_{H^{+}}$ &
$d_{H^{+}}$ & $w_{He^{+}}$ & $d_{He^{+}}$&$w_{MSE}$ \\ \hline
 & & & & & & & & \\
  SiVI 3S-3P  &50000.& 0.750E-02&-0.577E-04& 0.360E-04&-0.260E-04&0.687E-04&-0.500E-04&0.415E-02\\
  1226.7 A   &100000.& 0.532E-02&-0.739E-04&0.897E-04&-0.516E-04&0.173E-03&-0.103E-03&0.294E-02\\
 C= 0.11E+21 &200000.& 0.387E-02&-0.784E-04&0.177E-03&-0.920E-04&0.344E-03&-0.185E-03&0.212E-02\\
             &400000.&0.291E-02&-0.100E-03&0.269E-03&-0.139E-03&0.529E-03&-0.280E-03&0.169E-02\\
             &800000.&0.227E-02&-0.888E-04&0.347E-03&-0.189E-03&0.693E-03&-0.384E-03&0.147E-02\\
 & & & & & & & & \\
  SiVI 3S-3P  &50000. &0.681E-02&-0.660E-04 &0.346E-04&-0.294E-04 &0.660E-04&-0.566E-04&0.375E-02\\
  1187.2 A   &100000. &0.486E-02&-0.841E-04 &0.866E-04&-0.582E-04 &0.167E-03&-0.116E-03&0.265E-02\\
 C= 0.67E+20 &200000. &0.354E-02&-0.890E-04 &0.171E-03&-0.102E-03 &0.335E-03&-0.205E-03&0.191E-02\\
             &400000. &0.267E-02&-0.109E-03 &0.262E-03&-0.150E-03 &0.517E-03&-0.304E-03&0.152E-02\\
             &800000. &0.208E-02&-0.993E-04 &0.340E-03&-0.203E-03 &0.678E-03&-0.412E-03&0.132E-02\\
 & & & & & & & & \\
  SiVI 3S'-3P' &50000. &0.716E-02&-0.644E-04 &0.471E-04&-0.212E-04 &0.899E-04&-0.409E-04&0.389E-02\\
  1228.8 A   &100000. &0.509E-02&-0.599E-04 &0.113E-03&-0.424E-04 &0.219E-03&-0.842E-04&0.275E-02\\
 C= 0.72E+20 &200000. &0.370E-02&-0.615E-04 &0.213E-03&-0.771E-04 &0.416E-03&-0.155E-03&0.198E-02\\
             &400000. &0.278E-02&-0.770E-04 &0.314E-03&-0.119E-03 &0.618E-03&-0.241E-03&0.158E-02\\
             &800000. &0.216E-02&-0.691E-04 &0.389E-03&-0.164E-03 &0.774E-03&-0.331E-03&0.137E-02\\
 & & & & & & & & \\
  SiVI 3S'-3P' &50000. &0.572E-02&-0.376E-04 &0.399E-04&-0.131E-04 &0.762E-04&-0.252E-04&0.312E-02\\
  1087.4 A   &100000. &0.408E-02&-0.358E-04 &0.948E-04&-0.263E-04 &0.183E-03&-0.521E-04&0.220E-02\\
 C= 0.56E+20 &200000. &0.297E-02&-0.371E-04 &0.176E-03&-0.487E-04 &0.344E-03&-0.978E-04&0.158E-02\\
             &400000. &0.223E-02&-0.460E-04 &0.256E-03&-0.774E-04 &0.505E-03&-0.156E-03&0.126E-02\\
             &800000. &0.174E-02&-0.413E-04 &0.313E-03&-0.107E-03 &0.623E-03&-0.216E-03&0.109E-02\\
 & & & & & & & & \\
  SiVI 3S-3P  &50000. &0.842E-02&-0.107E-03 &0.361E-04&-0.374E-04 &0.690E-04&-0.721E-04&0.436E-02 \\
  1314.8 A   &100000. &0.598E-02&-0.118E-03 &0.927E-04&-0.740E-04 &0.179E-03&-0.147E-03&0.309E-02 \\
 C= 0.13E+21 &200000. &0.434E-02&-0.116E-03 &0.189E-03&-0.129E-03 &0.368E-03&-0.260E-03& 0.224E-02\\
             &400000. &0.326E-02&-0.148E-03 &0.295E-03&-0.190E-03 &0.582E-03&-0.384E-03&0.180E-02 \\
             &800000. &0.254E-02&-0.134E-03 &0.392E-03&-0.256E-03 &0.784E-03&-0.519E-03&0.155E-02 \\
 & & & & & & & & \\
  SiVI 3S-3P  &50000. &0.656E-02&-0.682E-04 &0.295E-04&-0.254E-04 &0.563E-04&-0.488E-04&0.343E-02 \\
  1145.4 A   &100000. &0.466E-02&-0.797E-04 &0.745E-04&-0.503E-04 &0.144E-03&-0.999E-04&0.242E-02 \\
 C= 0.10E+21 &200000. &0.338E-02&-0.787E-04 &0.149E-03&-0.887E-04 &0.290E-03&-0.178E-03&0.175E-02 \\
             &400000. &0.254E-02&-0.102E-03 &0.229E-03&-0.132E-03 &0.452E-03&-0.267E-03&0.141E-02 \\
             &800000. &0.198E-02&-0.909E-04 &0.300E-03&-0.179E-03 &0.599E-03&-0.364E-03&0.122E-02 \\
 & & & & & & & & \\
  SiVI 2P-3S  &50000. &0.209E-04&-0.851E-06 &0.885E-08 &0.227E-06 &0.169E-07 &0.438E-06&0.154E-04 \\
   100.2 A   &100000. &0.138E-04 &0.531E-06 &0.665E-07 &0.449E-06 &0.128E-06 &0.891E-06&0.109E-04 \\
 C= 0.73E+18 &200000. &0.976E-05 &0.815E-06 &0.288E-06 &0.780E-06 &0.571E-06 &0.157E-05&0.791E-05 \\
             &400000. &0.740E-05 &0.981E-06 &0.709E-06 &0.114E-05 &0.143E-05 &0.231E-05&0.628E-05 \\
             &800000. &0.577E-05 &0.918E-06 &0.131E-05 &0.153E-05 &0.262E-05 &0.311E-05&0.541E-05 \\
 & & & & & & & & \\
  SiVI 2P-3D  &50000. &0.267E-04&-0.113E-05 &0.147E-06&-0.328E-07 &0.280E-06&-0.631E-07&0.194E-04 \\
    83.8 A   &100000. &0.179E-04&-0.289E-06 &0.365E-06&-0.663E-07 &0.703E-06&-0.131E-06&0.137E-04 \\
 C= 0.77E+18 &200000. &0.130E-04&-0.490E-07 &0.715E-06&-0.128E-06 &0.139E-05&-0.257E-06&0.969E-05 \\
             &400000. &0.967E-05&-0.134E-06 &0.108E-05&-0.221E-06 &0.212E-05&-0.444E-06&0.747E-05 \\
             &800000. &0.741E-05&-0.802E-07 &0.139E-05&-0.321E-06 &0.277E-05&-0.647E-06&0.631E-05 \\
 & & & & & & & & \\
  SiVI 2P-3D  &50000. &0.266E-04&-0.116E-05 &0.147E-06&-0.165E-07 &0.280E-06&-0.318E-07& 0.193E-04\\
    83.8 A   &100000. &0.174E-04&-0.200E-06 &0.364E-06&-0.335E-07 &0.701E-06&-0.664E-07& 0.137E-04\\
 C= 0.35E+18 &200000. &0.125E-04 &0.533E-07 &0.712E-06&-0.662E-07 &0.139E-05&-0.132E-06& 0.967E-05\\
             &400000. &0.929E-05&-0.605E-08 &0.107E-05&-0.121E-06 &0.211E-05&-0.243E-06& 0.748E-05\\
             &800000. &0.709E-05 &0.492E-07 &0.139E-05&-0.188E-06 &0.275E-05&-0.380E-06& 0.631E-05\\
 & & & & & & & & \\ \hline
\end{tabular}
\end{center}
\end{table*}

\newpage
\addtocounter{table}{-1}
\begin{table*}
\caption{$Continued.$}
\label{tab:2}       
\begin{center}
\begin{tabular}{crrrrrrrr}
\hline transition  & T(K) & $w_{e}$ & $d_{e}$ & $w_{H^{+}}$ &
$d_{H^{+}}$ & $w_{He^{+}}$ & $d_{He^{+}}$&$w_{MSE}$ \\ \hline
 & & & & & & & \\
  SiVI 2P-3S'  &50000. &0.202E-04&-0.613E-06 &0.106E-07&0.247E-06 &0.202E-07 &0.475E-06&0.139E-04 \\
    96.7 A   &100000. &0.135E-04 &0.635E-06 &0.787E-07 &0.484E-06 &0.152E-06 &0.961E-06& 0.984E-05\\
 C= 0.45E+18 &200000. &0.958E-05 &0.879E-06 &0.317E-06 &0.831E-06 &0.629E-06 &0.167E-05& 0.712E-05\\
             &400000. &0.728E-05 &0.105E-05 &0.752E-06 &0.121E-05 &0.151E-05 &0.244E-05& 0.564E-05\\
             &800000. &0.570E-05 &0.993E-06 &0.139E-05 &0.158E-05 &0.278E-05 &0.321E-05& 0.486E-05\\
 & & & & & & & \\
  SiVI 2P-3D' &50000. &0.277E-04&-0.116E-05 &0.212E-06&-0.144E-07 &0.405E-06&-0.277E-07&0.178E-04\\
    81.2 A   &100000. &0.176E-04&-0.202E-06 &0.504E-06&-0.291E-07 &0.972E-06&-0.578E-07&0.126E-04\\
 C= 0.58E+18 &200000. &0.127E-04 &0.363E-07 &0.935E-06&-0.577E-07 &0.183E-05&-0.115E-06&0.902E-05\\
             &400000. &0.937E-05&-0.255E-07 &0.136E-05&-0.106E-06 &0.268E-05&-0.213E-06&0.710E-05\\
             &800000. &0.716E-05 &0.289E-07 &0.166E-05&-0.166E-06 &0.330E-05&-0.336E-06&0.612E-05\\
 & & & & & & & \\
  SiVI 2P-3D' &50000. &0.273E-04&-0.119E-05 &0.214E-06&-0.231E-07 &0.408E-06&-0.445E-07&0.179E-04 \\
    81.2 A   &100000. &0.174E-04&-0.260E-06 &0.508E-06&-0.468E-07 &0.981E-06&-0.928E-07&0.126E-04 \\
 C= 0.59E+18 &200000. &0.125E-04&-0.196E-07 &0.942E-06&-0.916E-07 &0.184E-05&-0.183E-06&0.904E-05 \\
             &400000. &0.923E-05&-0.845E-07 &0.137E-05&-0.161E-06 &0.270E-05&-0.324E-06&0.710E-05 \\
             &800000. &0.704E-05&-0.422E-07 &0.167E-05&-0.239E-06 &0.332E-05&-0.483E-06&0.610E-05 \\
 & & & & & & & \\
  SiVI 2P-3D' &50000. &0.270E-04&-0.107E-05 &0.219E-06&-0.148E-07 &0.419E-06&-0.285E-07&0.179E-04 \\
    81.0 A   &100000. &0.175E-04&-0.202E-06 &0.520E-06&-0.299E-07 &0.100E-05&-0.594E-07&0.127E-04 \\
 C= 0.60E+18 &200000. &0.126E-04 &0.317E-07 &0.960E-06&-0.592E-07 &0.187E-05&-0.118E-06&0.905E-05 \\
             &400000. &0.935E-05&-0.301E-07 &0.139E-05&-0.109E-06 &0.274E-05&-0.218E-06&0.711E-05 \\
             &800000. &0.714E-05 &0.195E-07 &0.169E-05&-0.170E-06 &0.336E-05&-0.343E-06&0.609E-05 \\
 & & & & & & & \\
  SiVI 3S'-3P'&50000. &0.439E-02&-0.460E-04 &0.328E-04&-0.168E-04 &0.626E-04&-0.324E-04&0.241E-02 \\
   918.8 A   &100000. &0.314E-02&-0.597E-04 &0.773E-04&-0.333E-04 &0.149E-03&-0.662E-04&0.171E-02\\
 C= 0.40E+20 &200000. &0.229E-02&-0.606E-04 &0.142E-03&-0.586E-04 &0.278E-03&-0.118E-03&0.123E-02 \\
             &400000. &0.173E-02&-0.737E-04 &0.206E-03&-0.869E-04 &0.407E-03&-0.176E-03&0.976E-03 \\
             &800000. &0.135E-02&-0.698E-04 &0.252E-03&-0.118E-03 &0.500E-03&-0.239E-03&0.845E-03 \\
 & & & & & & & \\
  Si VI 3S-3P &50000. &0.510E-02&-0.414E-04 &0.244E-04&-0.159E-04 &0.465E-04&-0.306E-04&0.263E-02 \\
   995.6 A   &100000. &0.362E-02&-0.514E-04 &0.604E-04&-0.317E-04 &0.116E-03&-0.630E-04&0.186E-02 \\
 C= 0.75E+20 &200000. &0.263E-02&-0.509E-04 &0.118E-03&-0.569E-04 &0.230E-03&-0.114E-03&0.134E-02 \\
             &400000. &0.198E-02&-0.655E-04 &0.179E-03&-0.864E-04 &0.352E-03&-0.175E-03&0.107E-02 \\
             &800000. &0.155E-02&-0.589E-04 &0.230E-03&-0.118E-03 &0.458E-03&-0.240E-03&0.923E-03 \\
 & & & & & & & \\\hline

\end{tabular}
\end{center}
\end{table*}

\bsp

\label{lastpage}

\end{document}